\documentclass[12pt]{article}
\usepackage{latexsym}
\usepackage{amsmath,amssymb,amsfonts,amsthm}
\usepackage[english]{babel}
\newcommand{\be}{\begin{equation}}
\newcommand{\ee}{\end{equation}}
\newcommand{\bea}{\begin{eqnarray}}
\newcommand{\eea}{\end{eqnarray}}
\newcommand{\nn}{\nonumber \\}
\newcommand{\p}[1]{(\ref{#1})}
\newcommand{\lb}{\label}

\begin{document}

\begin{titlepage}

\vspace*{1.5cm}

\begin{center}
\begin{Large}
{\Large\bf $\mathcal{N}=4$ Supersymmetric Landau Models}
\end{Large}
\vskip 0.3cm {\large {\sl }} \vskip 10.mm { V. Bychkov$^{\,1,\,2\,a}$, $\;\;\;$  E. Ivanov$^{\,2,\,b}$

}
\vspace{1cm}

{\it
$^{1}$ Department of General and Applied Physics, MIPT,\\
141700 Dolgoprudny, Moscow Region, Russia
\vspace{0.2cm}

$^2$  Bogoliubov Laboratory of Theoretical Physics, JINR, \\
141980 Dubna, Moscow Region, Russia
}
\end{center}
\vfill

\par
\begin{center}
{\bf ABSTRACT}
\end{center}
\begin{quote}
We present the first example of super Landau model with both ${\cal N}{=}4$ worldline supersymmetry and
non-trivial target space supersymmetry $ISU(2|2)$. The model also reveals a hidden second ${\cal N}{=}4$ supersymmetry which,
together with the manifest one, close on a worldline $SU(2|2)$.  We start from an off-shell action
in bi-harmonic ${\cal N}{=}4, d{=}1$ superspace and come to the component action with four bosonic and four fermionic
fields. Its bosonic core is the action of  generalized
$U(1)$ Landau model on $\mathbb{R}^4$ considered some time  ago by Elvang and
Polchinski. At each Landau level $N>0$ the wave functions are shown to form ``atypical''
($2N + 2N$)-dimensional multiplets of the worldline supergroup $SU(2|2)$. Some states have negative norms, but this trouble can be
evaded by redefining the inner product, like in other super Landau models. We promote the action to the most
general form compatible with off-shell ${\cal N}{=}4$ worldline supersymmetry and find the corresponding background
$U(1)$ gauge field to be generic self-dual on $\mathbb{R}^4\,$ and the target
superspace metric to remain flat.

\vfill \vfill \vfill \vfill \vfill \hrule width 5.cm \vskip 2.mm
{\small
\noindent $^a$ bychkov.vladimir@gmail.com\\
\noindent $^b$ eivanov@theor.jinr.ru\\
}
\end{quote}
\end{titlepage}
\setcounter{footnote}{0}


\numberwithin{equation}{section}

\section{Introduction}
After the pioneering paper \cite{0}, the name Landau
model is often used for any quantum-mechanical problem in which a
charged particle moves over some manifold in the background of an
external gauge field. Besides the original planar $2D$ Landau model,
with the particle moving on a plane under the influence of uniform
magnetic field orthogonal to the plane, much attention was paid to
some its curved generalizations, e.g. the Haldane model \cite{00},
where a charged particle moves on a two-dimensional sphere $S^2 \sim
SU(2)/U(1)$ in the field of magnetic monopole placed at the center,
as well as to some its higher-dimensional versions, with both
abelian and non-abelian gauge fields (see, e.g., \cite{zh,KN,8,KH}).
The Landau problem and its generalizations have a lot of
applications in various areas. In particular, they constitute a
theoretical basis of quantum Hall effect (QHE). Their most
characteristic feature is the presence of Landau Levels (LL) in the
energy spectrum, such that the gap between the ground state, i.e.
the Lowest Landau Level (LLL), and the excited LLs rapidly grows
with growth of the strength of the external gauge field. Thus in the
limit of strong external field only the LLL is relevant. In the
Lagrangian language, such a system is described by $d{=}1$ Wess-Zumino
(or Chern-Simons) action, with the phase space being a
non-commutative version of the original configuration space. This
intimate connection with non-commutative geometry was one of the
basic reasons of great revival of interest in Landau-type models
during the past decade.

Superextensions of the Landau model are models of non-relativistic particles moving
on supergroup manifolds. The study of such models can help to reveal possible manifestations
of supersymmetry in various versions of QHE (including the so called spin-QHE) and, perhaps, in other condensed matter systems.
From the mathematical point of view, superextended Landau models should bear a close relation
to the non-commutative supergeometry. It is also worth pointing out that sigma models with the supergroup
target spaces attract a lot of attention for the last years due to their intimate relation
to superbranes (see, e.g., \cite{MT,schom,Mi}). The super Landau models can be expected to follow from some of these
sigma models via dimensional reduction.

The Landau problems on the $(2|2)$-dimensional supersphere $SU(2|1)/U(1|1)$
and the $(2|4)$-dimensional superflag $SU(2|1)/[U(1)\times U(1)]$ as the simplest superextensions
of the $S^2$ Haldane model
were considered in \cite{1,2,SSph}. In order to better understand the common features of the super Landau models,
the planar limits of these models (with the curved target supermanifolds becoming the
$(2|2)$- and $(2|4)$-dimensional superplanes)
were also studied \cite{3,32,33,4}. They are superextensions of the original Landau model
and exhibit some surprising features.

First, the space of quantum states in these models involves ghosts, i.e. the states with negative norms,
which seemingly leads to
violation of unitarity.
The appearance of ghosts in $d{=}1$ supersymmetric models with second order kinetic terms for fermions
was earlier noticed by Volkov and Pashnev \cite{44}. The planar super Landau models suggest
a simple mechanism of evading the ghost problem. It was shown in \cite{33} that all norms of states in the
superplane models can be made non-negative at cost of introducing a proper metric operator in Hilbert space and so
redefining the inner product. There appear no difficulties with unitarity after such a redefinition.

The second feature closely
related to the one just mentioned is that the passing to the  new inner product (and so to the new definition
of hermitian conjugation)
makes manifest the hidden worldline ${\cal N}{=}2$ supersymmetry of the superplanar models, which so supply
examples of ${\cal N}{=}2$ supersymmetric quantum mechanics.

The presence of this worldline ${\cal N}{=}2$ supersymmetry was used
in \cite{43} to re-derive  the $(2|2)$ superplane Landau model from
the manifestly ${\cal N}{=}2$ supersymmetric worldline superfield
formalism. It was proposed there to construct new types  of
superextended Landau models, starting just from the superfield formalism, with the manifest
worldline supersymmetry as an input.
New ${\cal N}{=}2$ supersymmetric models were constructed in this way in \cite{4}.
They are generalizations of the superplane model
to the case with non-trivial target superspace metric and external gauge field.

In the present paper we apply the same approach to construct the first example of ${\cal N}{=}4$ supersymmetric Landau model
with a non-trivial target space supergroup. From the geometric point of view, it is a Landau-type model on
the flat $(4|4)$-dimensional target superspace $ISU(2|2)/SU(2|2)\,$ extending the Euclidean space $\mathbb{R}^4$,
with an extra worldline $\mathcal{N}{=}4$ supersymmetry.
Its off-shell action is formulated in terms of two superfields, bosonic and fermionic,
defined in the bi-harmonic ${\cal N}{=}4, d{=}1$ superspace \cite{6}.

The paper is organized as follows. In Sect.~2 we recall the salient feature of the ${\cal N}{=}2$ superfield construction of
the $(2|2)$ superplane Landau model \cite{43}. In Sect.~3 we give a short account of the bi-harmonic ${\cal N}{=}4, d{=}1$
superfield approach \cite{6}. The superfield and component actions of the new ${\cal N}{=}4$ super Landau model are constructed
in Sect.~4. The final action involves four bosonic and four fermionic $d{=}1$ fields and contains a coupling
to some external linear self-dual gauge field on $\mathbb{R}^4$.
We show that, besides the worldline $\mathcal{N}{=}4$
supersymmetry, the model has the target $ISU(2|2)$
symmetry.
The quantization is performed in Sect.~5. We show that at each Landau level $N$ the wave functions form the multiplets
of both target $ISU(2|2)$ and worldline ${\cal N}{=}4$ supersymmetries.
To avoid negative norms for some wave functions, it proves necessary to properly redefine
the inner product in the space of quantum states, like in the previously studied super Landau models \cite{33,4,SSph}.
Sect.~6 is devoted to the further analysis of the symmetry structure of the model. In particular, we find out
the existence of the second (on-shell)
worldline ${\cal N}{=}4$ supersymmetry, which, together with the first one, close on a worldline $SU(2|2)$ supersymmetry.
The wave functions form ``atypical'' multiplets of the latter.
In Sect.~7 we consider a generalization of the constructed ${\cal N}{=}4$ super Landau model
along the lines of ref. \cite{4}. The corresponding external bosonic gauge field proves to be generic self-dual
on $\mathbb{R}^4$,
while the target superspace metric remains flat, as opposed to the ${\cal N}{=}2$ models of ref. \cite{4}. Some problems
for the future study are outlined in the concluding Sect.~8. Appendices A and B contain some technical details.

\section{$ISU(1|1)$ super Landau model from $\mathcal{N}= 2, d=1$ superspace}
In this Section we remind some basic features of the manifestly $\mathcal{N}{=}2$ supersymmetric formulation of
the $(2|2)$ superplane Landau model \cite{43}.

\subsection{Superfield and component actions}
We start with the necessary definitions. The basic objects are two $\mathcal{N}{=}2, d{=}1$
chiral bosonic and fermionic superfields $\Phi$ and $\Psi$ of the same dimension.

The real $\mathcal{N}{=}2, d{=}1$ superspace is parametrized as:
\be
(\tau, \theta, \bar{\theta}).\lb{1}
\ee
The left- and right-handed chiral $d{=}1$ superspaces are defined as the coordinate sets
\be
(t_L, \theta), \quad (t_R, \bar{\theta}), \quad t_L = \tau + i\theta\bar{\theta}, t_R = \tau - i\theta\bar{\theta}.\lb{2}
\ee
It will be convenient to work in the left-chiral basis, so for brevity we will use the notation
$t_L \equiv t$, $t_R = t - 2i\theta\bar{\theta}$. In this basis, the $\mathcal{N}{=}2$
covariant derivatives are defined by
\be
\bar{D} = -\frac{\partial}{\partial\bar{\theta}}, \quad D = \frac{\partial}{\partial\theta}
+ 2i\bar{\theta}\partial_{t}, \quad \{D, \bar{D}\} = -2i\partial_t, \quad D^2 = \bar{D}^2 = 0.\lb{3}
\ee
The chiral superfields $\Phi$ and $\Psi$ obey the conditions
\be
\bar{D}\Phi = \bar{D}\Psi = 0\lb{4}
\ee
and, in the left-chiral basis, have the following component field contents:
\be
\Phi(t, \theta) = z(t) + \theta\chi(t), \quad \Psi(t, \theta) = \bar\zeta(t) + \theta h(t),\lb{5}
\ee
where the complex fields $z(t)$, $h(t)$ are bosonic and $\bar\zeta(t)$, $\chi(t)$ are fermionic.
The conjugated superfields, in the same basis, have the following $\theta$-expansions:
\be
\bar{\Phi} = \bar{z} - \bar{\theta}\bar\chi - 2i\theta\bar{\theta}\dot{\bar{z}}, \quad \bar{\Psi}
= \zeta + \bar{\theta}\bar{h} - 2i\theta\bar{\theta}\dot{\zeta}.\lb{6}
\ee
The superfield action yielding in components the superplane model action of ref. \cite{3,32,33} reads
\be
S = -\kappa\int dtd^2\theta\left(\Phi\bar{\Phi} + \Psi\bar{\Psi} + \frac{1}{2\sqrt{\kappa}}
[\Phi D\Psi - \bar\Phi\bar{D}\bar{\Psi}]\right)
= \int dt\,L.\lb{8}
\ee
Here $\kappa$ is a real parameter. The Berezin integral is normalized as
\be
\int d^2\theta(\theta\bar{\theta}) = 1.\lb{9}
\ee
After doing the Berezin integral, we find
\be
L = 2i\kappa(z\dot{\bar{z}} + \bar\zeta\dot{\zeta}) - \kappa(\chi\bar{\chi} + h\bar{h})
+ i\sqrt{\kappa}(z\dot{h} + \chi\dot{\bar{\zeta}} + \dot{\bar{z}}\bar{h} + \bar{\chi}\dot{\zeta}).\lb{10}
\ee
The fields $h$ and $\chi$ are auxiliary and can be eliminated by their equations of motion
\be
h = \frac{i}{\sqrt{\kappa}}\dot{\bar{z}}\,, \quad \chi = -\frac{i}{\sqrt{\kappa}}\dot{\zeta}\,.\lb{11}
\ee
Upon substituting this into the Lagrangian and integrating by parts, we obtain
\be
L = i \kappa(z\dot{\bar{z}} - \dot{z}\bar{z}+ \bar\zeta\dot{\zeta} - \dot{\bar{\zeta}}\zeta) +
(\dot{z}\dot{\bar{z}} + \dot{\zeta}\dot{\bar{\zeta}}).\lb{12}
\ee
This is the superplane model component Lagrangian \cite{3,33}.
By construction, the superfield action \p{8} is $\mathcal{N}{=}2$ supersymmetric, so are the component
Lagrangians \p{10} and \p{12}.
The $\mathcal{N}{=}2$ transformations of the component fields can be found from
\be
\delta\Phi = -[\epsilon Q - \bar{\epsilon}\bar{Q}]\Phi, \quad \delta\Psi = -[\epsilon Q - \bar{\epsilon}\bar{Q}]\Psi,\lb{15}
\ee
where, in the left-chiral basis,
\be
Q = \frac{\partial}{\partial\theta}, \quad \bar{Q} = -\frac{\partial}{\partial\bar{\theta}} + 2i\theta\partial_t,
\quad \{Q, \bar{Q}\} = 2i\partial_t\,.\lb{16}
\ee
It follows from \p{15}, \p{16} that, off shell,
\be
\delta z = -\epsilon\chi, \quad \delta\chi = -2i\bar{\epsilon}\dot{z}, \quad \delta\zeta= -\bar\epsilon \,\bar h,
\quad \delta h = -2i\bar{\epsilon}\dot{\bar{\zeta}}.\lb{17}
\ee
With the on-shell values \p{11} for the auxiliary fields, these transformations become
\be
\delta z = \frac{i}{\sqrt{\kappa}}\epsilon\dot{\zeta}, \quad \delta\zeta = \frac{i}{\sqrt{\kappa}}\bar{\epsilon}\dot{z}.\lb{18}
\ee
These are basically the same transformation laws as those found in \cite{4} (up to rescaling of $\epsilon, \bar{\epsilon}$).
As usual, they close on $t$-translations only with making use of the equations of motion for physical fields, while
\p{17} close without any help from the equations of motion.

It is worth pointing out that the ${\cal N}{=}2$ superfield formulation of the superplane Landau model described above is well
defined only at $\kappa \neq 0$.

\subsection{$ISU(1|1)$ symmetry}
Besides $\mathcal{N}{=}2$ supersymmetry, the superplane model also possesses the target space graded $ISU(1|1)$ symmetry.

The inhomogeneous translation part of this internal  supersymmetry acts as constant shifts of superfield:
\be
\delta\Phi = b, \quad \delta\Psi = \bar\nu\,,
\ee
where $b$ and $\nu$ are even and odd complex parameters. They just produce shifts of the fields $z$ and $\zeta$
\be
\delta z = b\,, \quad \delta \zeta = \nu\,.\lb{tarstran}
\ee
The fermionic transformations of the homogeneous $SU(1|1)$ part are realized as
\be
\delta\Phi = \bar{D}\Big(\omega\bar{\theta}\bar{\Psi} - \frac{1}{2\sqrt{\kappa}}\bar{\omega}\theta D\Phi\Big),
\quad \delta\Psi = \bar{D}\Big(\omega\bar{\theta}\bar{\Phi} - \frac{1}{2\sqrt{\kappa}}\bar{\omega}\theta D\Psi\Big),
\ee
where $\omega$ and $\bar{\omega}$ are the relevant complex Grassmann parameters. They close on
the bosonic $U(1)$ transformations
\bea
\delta \Phi = i\alpha\left\{ \Phi - \bar D\Big[\theta\bar\theta \Big(D\Phi
+ \frac{i}{\sqrt{\kappa}}\dot{\bar{\Psi}}\Big)\Big]\right\}, \;
\delta \Psi = -i\alpha \left\{\Psi -\bar D\Big[\theta\bar\theta\Big(D\Psi
- \frac{i}{\sqrt{\kappa}}\dot{\bar{\Phi}} \Big)\Big] \right\}.
\eea

Though these superfield $SU(1|1)$ rotations look rather cumbersome, they give rise to the very simple off-shell
transformations of the physical fields $z$ and $\zeta$:
\be
\delta\left(
   \begin{array}{c}
     z\\
     \zeta
   \end{array}
   \right) = \left(
     \begin{array}{cc}
       i\alpha&  {\omega}\\
       \bar\omega& i\alpha
     \end{array}
   \right)
   \left(
   \begin{array}{c}
     z\\
     \zeta
   \end{array}
   \right). \lb{tarsrot}
\ee
The transformations of the auxiliary fields are
$$
\delta_\omega \chi = -\frac{i}{\sqrt{\kappa}}\,\bar\omega\, \dot z\,, \; \delta_\omega h =
-\frac{i}{\sqrt{\kappa}}\,\bar\omega\, \dot{\bar\zeta}\,, \quad
\delta_\alpha \chi = \frac{1}{\sqrt{\kappa}}\,\alpha\, \dot \zeta\,,\; \delta_\alpha h
= \frac{1}{\sqrt{\kappa}}\,\alpha\,\dot{\bar z}\,.
$$
They are consistent with the on-shell expressions \p{11}.

In quantum theory, the generators associated with the target supertranslations \p{tarstran} and super-rotations \p{tarsrot} are
given by the expressions \cite{3,33}
\be
P_z = -i(\partial_z + k\bar{z}), \quad P_{\bar{z}} = -i(\partial_{\bar{z}} - kz), \quad \Pi_{\zeta} =
\partial_{\zeta} + k\bar{\zeta}, \quad \Pi_{\bar{\zeta}} = \partial_{\bar{\zeta}} + k\zeta
\ee
and
\be
Q = z\partial_{\zeta} - \bar{\zeta}\partial_{\bar{z}}, \quad \bar{Q} = \bar{z}\partial_{\bar{\zeta}} + \zeta\partial_z,
\quad C = z\partial_z + \zeta\partial_{\zeta} - \bar{z}\partial_{\bar{z}} - \bar{\zeta}\partial_{\bar{\zeta}}.
\ee
These generators form the algebra of the supergroup $ISU(1|1)$
\be
(P_z, P_{\bar{z}}, \Pi_{\zeta}, \Pi_{\bar{\zeta}}) \rtimes SU(1|1) = ISU(1|1).
\ee
Note that the parameter $\kappa$ plays the role of central charge in the quantum algebra of supertranslations:
\be
[P_z, P_{\bar{z}}] = 2\kappa\,, \quad \{\Pi_\zeta, \Pi_{\bar\zeta}\} = 2\kappa\,.
\ee

The structure of the space of quantum states of the ${\cal N}{=}2$ super Landau model, the realization
of various symmetry generators in it,
as well as the explicit form of the metric operator making norms of all states positive-definite can be found in \cite{33}.

\setcounter{equation}{0}
\section{Bi-harmonic ${\cal N}=4$ superspace: basic notions}
Our aim will be to construct a generalization of the ${\cal N}{=}2$ model of the previous Section, such that it possesses
the worldline ${\cal N}{=}4$ supersymmetry. Such an extension is not unique; leaving the study of all possible versions of it
for the future,
here we will do this by extending the previously used bosonic and fermionic chiral $({\bf 2,2,0})$ and $({\bf 0, 2, 2})$
multiplets to the
$({\bf 4,4,0})$ and $({\bf 0,4,4})$ multiplets of ${\cal N}{=}4$ supersymmetry\footnote{The symbol (${\bf n_1, n_2, n_3}$)
denotes an off-shell multiplet with ${\bf n_1}$ physical bosonic fields,
${\bf n_2}$ fermionic fields and ${\bf n_3 = n_2 - n_1}$ additional bosonic fields.}.
 It turns out that these bosonic and fermionic
${\cal N}{=}4$ multiplets should be ``mirror'' (or ``twisted'') to each other. The natural framework
for a simultaneous description
of these two different sorts of ${\cal N}{=}4$ multiplets is provided by the bi-harmonic
${\cal N}{=}4, d{=}1$ superspace \cite{6}
which is an extension of the more familiar harmonic superspace involving one set
of $SU(2)$ harmonic variables \cite{HSS,7,5}. Here
we briefly outline this universal approach.

We begin with the ordinary $\mathcal{N}{=}4$, $d{=}1$ superspace in the notation with  both $SU(2)$ automorphism
groups being manifest. It is defined as the set of coordinates
\be
z := (t, \theta^{ia}),\lb{2.1}
\ee
in which $\mathcal{N}{=}4$, $d{=}1$ supersymmetry is realized by means of the transformations
\be
\delta t = -i\epsilon^{ia}\theta_{ia}, \quad \delta\theta^{ia} = \epsilon^{ia}.\lb{st}
\ee
The Grassmann coordinate $\theta^{ia}$ (as well as the parameters $\epsilon^{ia}$) form a real quartet
of the full automorphism group $SO(4) \sim SU(2)_L\times SU(2)_R$, $\overline{(\theta^{ia})}
= \theta_{ia} = \epsilon_{ik}\epsilon_{ab}\theta^{kb}$. The indices $i$ and $a$ are doublet indices of the left
and right $SU(2)$ automorphism groups, respectively. The corresponding covariant spinor derivatives are defined as
\be
D_{ia} = \frac{\partial}{\partial\theta^{ia}} + i\theta_{ia}\partial_t, \quad \bar{D}^{ia}
= -\frac{\partial}{\partial\theta_{ia}} - i\theta^{ia}\partial_t = -\epsilon^{ik}\epsilon^{ab}D_{kb},
\ee
\be
\{D_{ia}, D_{kb}\} = 2i\epsilon_{ik}\epsilon_{ab}\partial_t.
\ee
In the central basis, the $\mathcal{N}{=}4$, $d{=}1$ bi-harmonic superspace (bi-HSS) is defined
as the following extension of \p{2.1}
\be
(z, u, v) := (t, \theta^{ia}, u_i^{\pm1}, v_b^{\pm1}).\lb{cb}
\ee
Here $u_i^{\pm1} \in SU(2)_L/U(1)_L$ and $v_a^{\pm1} \in SU(2)_R/U(1)_R$ are two independent
sets of $SU(2)$ harmonic variables.
The harmonics $u_i^{\pm1}$ satisfy the standard relations \cite{HSS,7}
\be
u_i^{-1} = \overline{(u^{1i})}, u^{1i}u_i^{-1} = 1 \Leftrightarrow u_i^1u_k^{-1} - u_k^1u_i^{-1} = \epsilon_{ik}.
\ee
The same relations are valid for $v_a^{\pm1}$, with the change $i, k \rightarrow a, b$.

A specific feature of the $\mathcal{N}{=}4$, $d{=}1$ bi-HSS is the existence of two types of analytic
bases with the analytic subspaces including half of the Grassmann variables, as compared to the full Grassmann
dimension four of bi-HSS. These two analytic bases are spanned by the following coordinate sets
\be
(z_{+}, u, v) := (t_{+} = t + i(\theta^{1,1}\theta^{-1,-1} + \theta^{-1,1}\theta^{1,-1})\,, \theta^{1,1}\,,
\theta^{1,-1}, \theta^{-1,1}, \theta^{-1,-1}, u_{i}^{\pm 1}, v_{a}^{\pm 1})\,,\lb{anb1}
\ee
\be
(z_{-}, u, v) := (t_{-} = t + i(\theta^{1,1}\theta^{-1,-1} - \theta^{-1,1}\theta^{1,-1})\,,
\theta^{1,1}, \theta^{1,-1}, \theta^{-1,1}, \theta^{-1,-1}, u_{i}^{\pm 1}, v_{a}^{\pm 1})\,,\lb{anb2}
\ee
where
\be
\theta^{m,n} := \theta^{ia}u_i^mv_a^n, \quad m,n = \pm1\,.
\ee
Defining harmonic projections of the spinor derivatives as
\be
D^{m,n} = D^{ia}u_i^mv_a^n\,,
\ee
\be
(D^{1,1})^2 = (D^{1,-1})^2 = (D^{-1,1})^2 = (D^{-1,-1})^2 = \{D^{\pm1,1}, D^{\pm1,-1}\}
= \{D^{1,\pm1}, D^{-1,\pm1}\} = 0\,,
\ee
\be
\{D^{1,1}, D^{-1,-1}\} = -\{D^{1,-1}, D^{-1,1}\} = 2i\partial_t\,,
\ee
it is easy to show that, in the above bases, they have the form
\be
D^{1,1} = \frac{\partial}{\partial\theta^{-1,-1}}\,, \quad D^{1,-1} = -\frac{\partial}{\partial\theta^{-1,1}}\,, \nonumber
\ee
\be
D^{-1,1} = -\frac{\partial}{\partial\theta^{1,-1}} + 2i\theta^{-1,1}\partial_{t_+}\,, \quad
D^{-1,-1} = \frac{\partial}{\partial\theta^{1,1}} + 2i\theta^{-1,-1}\partial_{t_+}\,,
\ee
and
\be
D^{1,1} = \frac{\partial}{\partial\theta^{-1,-1}}\,, \quad D^{-1,1} = -\frac{\partial}{\partial\theta^{1,-1}}\,, \nonumber
\ee
\be
D^{1,-1} = -\frac{\partial}{\partial\theta^{-1,1}} + 2i\theta^{1,-1}\partial_{t_-}, \quad D^{-1,-1}
= \frac{\partial}{\partial\theta^{1,1}} + 2i\theta^{-1,-1}\partial_{t_-}\,.
\ee
The fact that two different pairs of covariant spinor derivatives are reduced
to the partial derivatives in these bases implies the existence of two analytic subspaces
which are closed under the full $\mathcal{N}{=}4$ supersymmetry. Hence there are two sorts
of analytic superfields defined as unconstrained functions on these analytical superspaces:
\be
(\zeta_{+},u,v) := (t_{+}, \theta^{1,1}, \theta^{1,-1}, u_{i}^{\pm1}, v_{a}^{\pm1}),\lb{anss1}
\ee
\be
D^{1,1}\Phi_{(+)} = D^{1,-1}\Phi_{(+)} = 0 \;\Rightarrow \; \Phi_{(+)} = \phi_{(+)}(\zeta_{+},u,v)\,,
\ee
and
\be
(\zeta_{-},u,v) := (t_{-}, \theta^{1,1}, \theta^{-1,1}, u_{i}^{\pm1}, v_{a}^{\pm1})\,, \lb{anss2}
\ee
\be
D^{1,1}\Phi_{(-)} = D^{-1,1}\Phi_{(-)} = 0 \;\Rightarrow \; \Phi_{(-)} = \phi_{(-)}(\zeta_{-},u,v)\,.
\ee
The analytic superspaces are real with respect to some generalized $\sim$
conjugation the implementation of which on coordinates can be found in \cite{6}. As a consequence, one can impose proper
reality conditions on the analytic superfields.

In the harmonic superspace approach,  harmonic derivatives play an important role.
The harmonic derivatives with respect to harmonics $u_i^{\pm1}$ and $v_a^{\pm1}$ in the central basis
are defined as
\be
\partial^{\pm2,0} = u_i^{\pm1}\frac{\partial}{\partial u_i^{\mp1}}\,, \quad \partial^{0,\pm2}
= v_a^{\pm1}\frac{\partial}{\partial v_a^{\mp1}}\,,
\ee
\be
\partial_u^0 = u_i^1\frac{\partial}{\partial u_i^1} - u_i^{-1}\frac{\partial}{\partial u_i^{-1}}\,,
\quad \partial_v^0 = v_a^1\frac{\partial}{\partial v_a^1} - v_a^{-1}\frac{\partial}{\partial v_a^{-1}}\,.
\ee
These sets form two mutually commuting $SU(2)$ algebras
\be
[\partial^{2,0}, \partial^{-2,0}] = \partial_u^0\,, \quad [\partial_u^0, \partial^{\pm2,0}] = \pm2\partial^{\pm2,0}\,,\lb{com1}
\ee
\be
[\partial^{0,2}, \partial^{0,-2}] = \partial_v^0\,, \quad [\partial_v^0, \partial^{0,\pm2}] = \pm2\partial^{0,\pm2}\,.\lb{com2}
\ee
In the analytic bases \p{anb1} and \p{anb2} the harmonic derivatives acquire additional terms. For example, in the basis \p{anb1}:
\be
D^{\pm2,0} = \partial^{\pm2,0} \pm 2i\theta^{\pm1,\pm1}\theta^{\pm1,\mp1}\partial_{t_+}
+ \theta^{\pm1,\pm1}\frac{\partial}{\partial\theta^{\mp1,\pm1}} + \theta^{\pm1,\mp1}\frac{\partial}{\partial\theta^{\mp1,\mp1}}\,,
\ee
\be
D^{0,\pm2} = \partial^{0,\pm2} + \theta^{\pm1,\pm1}\frac{\partial}{\partial\theta^{\pm1,\mp1}}
+ \theta^{\mp1,\pm1}\frac{\partial}{\partial\theta^{\mp1,\mp1}}\,,
\ee
\be
D_u^0 = \partial_u^0 + \Big(\theta^{1,1}\frac{\partial}{\partial\theta^{1,1}}
+ \theta^{1,-1}\frac{\partial}{\partial\theta^{1,-1}}
- \theta^{-1,1}\frac{\partial}{\partial\theta^{-1,1}} - \theta^{-1,-1}\frac{\partial}{\partial\theta^{-1,-1}}\Big),
\ee
\be
D_v^0 = \partial_v^0 + \Big(\theta^{1,1}\frac{\partial}{\partial\theta^{1,1}}
+ \theta^{-1,1}\frac{\partial}{\partial\theta^{-1,1}}
- \theta^{1,-1}\frac{\partial}{\partial\theta^{1,-1}} - \theta^{-1,-1}\frac{\partial}{\partial\theta^{-1,-1}}\Big).
\ee
Their commutation relations, being basis-independent, are given by the same formulas \p{com1} and \p{com2}.

Let us define integration measures on the full $\mathcal{N}{=}4$, $d{=}1$ bi-HSS and on its analytic subspaces:
\be
\mbox{Full bi-HSS}: \qquad \int\mu := \int dtdudv(D^{-1,-1}D^{-1,1}D^{1,1}D^{1,-1}),
\ee
\be
\mbox{Analytic bi-HSS 1}: \quad \int\mu^{(-2,0)}_{A+} := \int dt_+dudv(D^{-1,-1}D^{-1,1}),
\ee
\be
\mbox{Analytic bi-HSS 2}: \quad \int\mu^{(0,-2)}_{A-} := \int dt_-dudv(D^{-1,-1}D^{1,-1})\,.
\ee
They are normalized in such a way that
\be
\int\mu(\theta^{-1,-1}\theta^{-1,1}\theta^{1,1}\theta^{1,-1})\times... = \int dtdudv\times...,
\ee
\be
\int\mu^{(-2,0)}_{A+}(\theta^{1,1}\theta^{1,-1})\times... = \int dt_{+}dudv\times...,
\ee
\be
\int\mu^{(0,-2)}_{A-}(\theta^{1,1}\theta^{-1,1})\times... = \int dt_{-}dudv\times...
\ee

Finally, it is worth recalling the rules of integration over harmonic variables.
Symmetric monomials constructed from $u^{\pm 1}_i\,$,
\be
(u^1)^{(m}(u^{-1})^{n)} \equiv u^{1(i_1}...u^{1i_m}u^{-1j_1}...u^{-1j_n)}\,,
\ee
form orthogonal basis in the space of the functions on the 2-sphere $S^2$:
\be
\int du (u^1)^{(m}(u^{-1})^{n)}(u^1)_{(k}(u^{-1})_{l)}
= \frac{(-1)^nm!n!}{(m+n+1)!}\delta^{(i_1}_{(j_1}...\delta^{i_{n+m})}_{j_{n+m})}\delta_{ml}\delta_{nk}.
\ee
In what follows we shall need only the special case of this formula
\be
\int du \; u^1_iu^{-1}_j = \frac{1}{2}\epsilon_{ij}.
\ee
The similar relations hold for $v^{\pm 1}_a$.

The properties and relations quoted here are sufficient for
construction of $\mathcal{N}{=}4$ supersymmetric Landau model.

\setcounter{equation}{0}
\section{Landau model with $\mathcal{N}=4$ supersymmetry}

\subsection{Superfield and component actions}
Superfields $q^{1,0A}$ and $\psi^{0,1B}$ (A,B = 1,2) will be the
basic elements of our $\mathcal{N}{=}4$ supersymmetric Landau model.
These superfields are, respectively, bosonic and fermionic,  and
they have the fields contents $({\bf 4,4,0})$ and $({\bf
0,4,4})$\footnote{Division of these sets into physical and auxiliary
fields depends on the choice of the invariant action. Like in the
${\cal N}{=}2$ case, the fermionic fields in $q^{1,0A}$ and the additional
bosonic fields in $\psi^{0,1B}$ will be auxiliary, while the rest of
fields will be physical. This deviation from the standard divisions
of such multiplets into the physical and auxiliary subsets is of
course related to the fact that $q^{1,0A}$ and $\psi^{0,1B}$ have
the same dimension, which is necessary for realizing on them some
internal supersymmetry generalizing $ISU(1|1)$ symmetry of the
${\cal N}{=}2$ case. Correspondingly, physical bosons and fermions
will enter the component action on equal footing, with the
second-order kinetic terms.}. We impose on them the following
analytic and harmonic constraints \be (a)\quad D^{1,1}q^{1,0A} =
D^{1,-1}q^{1,0A} = 0, \qquad (b)\quad D^{2,0}q^{1,0A} =
D^{0,2}q^{1,0A} = 0, \lb{con_a} \ee and \be (a)\quad
D^{1,1}\psi^{0,1A} = D^{-1,1}\psi^{0,1A} = 0, \qquad (b)\quad
D^{2,0}\psi^{0,1A} = D^{0,2}\psi^{0,1A} = 0. \lb{con_b} \ee
The first
conditions in \p{con_a} and \p{con_b} tell us that the superfields
$q^{1,0A}$ and $\psi^{0,1B}$ ``live'' on the analytic subspaces
$(\zeta_+, u, v)$ and $(\zeta_-, u, v)\,$, respectively. Taking into
account the reality conditions ($\widetilde{q^{1,0A}} =
\epsilon_{AB}q^{1,0B}$) we can then solve the condition
(\ref{con_a}b) and obtain the following final component expansion
for the superfield $q^{1,0A}$: \be q^{1,0A} = f^{iA}(t_+)u_{i}^{1} +
\psi^{aA}(t_+)v_{a}^{-1}\theta^{1,1} -
\psi^{aA}(t_+)v_{a}^{1}\theta^{1,-1} -
2i\dot{f}^{iA}(t_+)u_{i}^{-1}\theta^{1,1}\theta^{1,-1}.\lb{sq} \ee
Similarly, the component expansion for the superfield $\psi^{0,1A}$
($\widetilde{\psi^{0,1A}} = \epsilon_{AB}\psi^{0,1B}$) reads: \be
\psi^{0,1A} = \chi^{aA}(t_-)v_{a}^{1} +
h^{iA}(t_-)u_{i}^{1}\theta^{-1,1} -
h^{iA}(t_-)u_{i}^{-1}\theta^{1,1} -
2i\dot{\chi}^{aA}(t_-)v_{a}^{-1}\theta^{1,1}\theta^{-1,1}.\lb{spsi}
\ee

In order to construct $\mathcal{N}{=}4$ supersymmetric Landau model action, we need one more object, namely,
the superfield $V^{1,0A} = D^{1,-1}\psi^{0,1A}$. It is easy to show that $V^{1,0A}$ ``live''
on the subspace $(\zeta_+, u, v)$,
since $D^{1,1}V^{1,0A} = D^{1,-1}V^{1,0A} = 0$. We obtain:
\be
V^{1,0A} = -h^{iA}(t_{+})u_{i}^{1} + 2i\dot{\chi}^{aA}(t_{+})v_{a}^{-1}\theta^{1,1} - 2i\dot{\chi}^{aA}(t_{+})v_{a}^{1}\theta^{1,-1}
+ 2i\dot{h}^{iA}(t_{+})u_{i}^{-1}\theta^{1,1}\theta^{1,-1}.
\ee
The fields $f^{iA}(t_+)$, $h^{iA}(t_-)$ are real  bosonic, while the fields $\psi^{aA}(t_+)$, $\chi^{aA}(t_-)$
are real fermionic.
The reality conditions are as follows
\be
\overline{f^{iA}} = \epsilon_{ij}\epsilon_{AB}f^{jB}, \quad \overline{h^{iA}} = \epsilon_{ij}\epsilon_{AB}h^{jB},
\quad \overline{\psi^{aA}} = \epsilon_{ab}\epsilon_{AB}\psi^{bB}, \quad \overline{\chi^{aA}}
= \epsilon_{ab}\epsilon_{AB}\chi^{bB}.
\ee

Now we can construct the superfield action for $\mathcal{N}{=}4$ supersymmetric  $(4|4)$ Landau model as a natural
generalization of the ${\cal N}{=}2$ action \p{8}
\be
S = \frac{\kappa}{2i}\Big(\int\mu^{-2,0}q^{1,0A}q^{1,0B}C_{AB} - i\int\mu^{0,-2}\psi^{0,1A}\psi^{0,1B}\epsilon_{AB}
+ \frac{1}{\sqrt{\kappa}}\int\mu^{-2,0}q^{1,0A}D^{1,-1}\psi^{0,1B}\epsilon_{AB}\Big). \lb{supact}
\ee
Here, $C_{AB}$ and $\epsilon_{AB}$ are symmetric and standard skew-symmetric constant tensors, respectively,
$\kappa \neq 0$ is a constant. Without loss of generality, we can choose
\be
C^{AB}C_{AB} = 2. \lb{C2}
\ee
In fact, we started from the more general action, with some arbitrary coefficients and constant matrices
before the three terms
in \p{supact}, and found that it can be reduced to the form \p{supact} with the condition \p{C2}
after some redefinitions
of the involved superfields.

The first two terms in \p{supact} are direct analogs of the first two terms in \p{8}, while the third interaction
term is an analog of the third term in \p{8}. It is important to point out that the necessity to use the ``mirror''
fermionic superfield $\psi^{0,1A}$ comes just as a necessary condition for constructing this interaction term.
A simple analysis
shows that no such bilinear interaction terms can be constructed from the bosonic and fermionic superfields of the
same harmonic analyticity. Also note that this mixed term admits a ``dual'' representation as an integral over another
analytic subspace:
$$
\sim \int \mu^{0,-2}\tilde{V}^{0,1A}\psi^{0,1 B}\rho \epsilon_{AB}\,, \quad \tilde{V}^{0,1A} = D^{-1, 1}q^{1,0 A}\,.
$$

Now we are prepared to derive the component Lagrangian of the model by performing integration
over Grassmann and harmonic variables.
We obtain:
\be
L = \frac{\kappa}{2i}\Big[(2i\dot{f}^{iA}f_{i}^{B} - \psi^{aA}\psi_{a}^{B})C_{AB} + (2\dot{\chi}^{aA}\chi_{a}^{B}
-i h^{iA}h_{i}^{B})\epsilon_{AB} - \frac{2i}{\sqrt{\kappa}}(\dot{f}^{iA}h_{i}^{B}
+ \psi^{aA}\dot{\chi}_{a}^{B})\epsilon_{AB}\Big].\lb{Lhpsi}
\ee
The fields $h^{iA}$ and $\psi^{aA}$ enter this Lagrangian without time derivatives and, hence, are auxiliary fields.
The remaining $(4 + 4)$ fields $f^{ia}$ and $\chi^{a A}$ are physical.
After eliminating the auxiliary fields by their algebraic equations of motion,
\be
\frac{\partial L}{\partial h^{iA}} = 0 \Rightarrow h_{iA} = -\frac{1}{\sqrt{\kappa}}\,\dot{f}_{iA},\lb{h}
\ee
\be
\frac{\partial L}{\partial \psi^{aA}} = 0 \Rightarrow \psi_{aA} = \frac{i}{\sqrt{\kappa}}\,C_{AB}\dot{\chi}_a^B\,,\lb{psi}
\ee
and substituting these expressions back into \p{Lhpsi}, we obtain
\be
L = \kappa C_{AB}\dot{f}^{iA}f_{i}^{B} - i\kappa\dot{\chi}^{aA}\chi_{aA} + \frac{1}{2}\left(\dot{f}^{iA}\dot{f}_{iA}
+ iC_{AB}\dot{\chi}^{aA}\dot{\chi}_{a}^{B}\right).\lb{Lfchi}
\ee

This Lagrangian is the sought ${\cal N}{=}4$ supersymmetric extension of the ${\cal N}{=}2$ Landau model Lagrangian \p{10}.
It includes
four real bosonic fields, so what we obtained is a superextension of the bosonic Landau-type model,
in which a particle moves over
four-dimensional Euclidean space $\mathbb{R}^4$ in an external U(1) gauge field.
This coupling is provided just by the first term in \p{Lfchi}. It can be rewritten as
\be
{\cal A}_{iB}\dot{f}^{iB}\,, \quad {\cal A}_{iB} = -\kappa C_B^{\;\;D}f_{i D}. \lb{LF}
\ee
Defining the covariant field strength,
$$
{\cal F}_{iA\,jB} = \partial_{iA}{\cal A}_{jB} - \partial_{jB}{\cal A}_{iA}\,,
$$
one finds
\be
{\cal F}_{iA\,jB} = -2\kappa\,C_{AB}\epsilon_{ij}\,. \lb{strength}
\ee
This means that the external Maxwell field is necessarily {\it self-dual}:
\be
{\cal F}_{(iA\,j)B} = 0 \lb{selfd}
\ee
(brackets $(\;)$ mean symmetrization with respect to the indices $i, j$). Thus the external Maxwell field should
be self-dual, as distinct from an unconstrained field strength $\sim \kappa$ in $2D$ case.
As we shall see, this self-duality of the external gauge field is necessarily implied
by the underlying ${\cal N}{=}4$ worldline
supersymmetry, like in conventional ${\cal N}{=}4$ mechanics models
(with the first-order kinetic terms for fermions) \cite{Pol,5,AW,SKon,IKS,DI}.

For the purposes of quantization it will be convenient to make one more simplification. It is related to the presence
of three $SU(2)$ groups in \p{supact} and \p{Lhpsi}.
While the automorphism $SU(2)_{L,R}$ symmetries acting on the indices $i$ ad $a$ of the component
fields are unbroken,
one more SU(2)$_{ext}$ acting on the capital doublet indices is necessarily
broken by the first term in \p{supact}, which includes the constant symmetric
tensor $C_{AB}$\footnote{This implies the breaking of the
``Lorentz'' $SO(4)\sim SU(2)_L\times SU(2)_{ext}$ symmetry of $\mathbb{R}^4$ down to $SU(2)_L\times U(1)_{ext}\,$.}.
In what follows,
without loss of generality, we can make use of this broken $SU(2)_{ext}$ to bring $C^{AB}$ into the particular form
\be
C^{12} = i\,,
\ee
with all other components vanishing.

Using all these simplifications, we rewrite Lagrangian \p{Lfchi} in a different notation, by passing
from the  quartets $f^{iA}$ and $\chi^{aA}$ to the doublets of complex fields $z, u, \zeta$ and $\xi$
\be
z = f^{11}, \quad \bar{z} = f^{22}, u = f^{21}, \quad \bar{u} = -f^{12},\lb{redef1}
\ee
\be
\zeta = \chi^{11}, \quad \bar{\zeta} = \chi^{22}, \quad \xi = \chi^{21}, \quad \bar{\xi} = -\chi^{12}. \lb{redef2}
\ee
Then
\be
L = |\dot{z}|^2 + |\dot{u}|^2 - i\kappa(\dot{z}\bar{z}-\dot{\bar{z}}z + \dot{u}\bar{u}-\dot{\bar{u}}u)
+ \dot{\zeta}\dot{\bar{\zeta}} + \dot{\xi}\dot{\bar{\xi}} - i\kappa(\dot{\zeta}\bar{\zeta}
+ \dot{\bar{\zeta}}\zeta + \dot{\xi}\bar{\xi} + \dot{\bar{\xi}}\xi).\lb{Lz}
\ee

Thus, using the superfield approach, we derived the component Lagrangian \p{Lz} for $\mathcal{N}{=}4$ extended supersymmetric
Landau model, where the worldline ${\cal N}{=}4$ supersymmetry is built-in by construction.
Though the action \p{Lz} is a sum of two copies of the ${\cal N}{=}2$ Landau model actions \p{12},
it possesses a rich symmetry structure,
as will be demonstrated in the next Sections. Its bosonic sector is just the action of four-dimensional
$U(1)$ Landau-type model discussed
in \cite{8}. The Lorentz-force term \p{LF} is rewritten as
\be
{\cal A}_{i}\dot{z}^i + \bar{\cal A}^i\dot{\bar z}_{i}\,,
\ee
where $z^i \equiv (z, \; u), \bar z_{i} \equiv (\bar z, \; \bar u)$ and
\be
{\cal A}_{i} = -i\kappa\, \bar z_i\,, \quad  \bar{\cal A}^{i} = i\kappa\,z^i\,.
\ee
One can check that the components of the background gauge field in this $SU(2)_L$ covariant notation are given by
\be
{\cal F}_{i}^{\;l} = 2i\kappa\,\delta_{i}^l\,, \quad {\cal F}_{i l}
= \bar{\cal F}^{il} =0\,,
\ee
which coincide with those in \cite{8}.

\subsection{Worldline supersymmetry}
In this subsection, we give how $\mathcal{N}{=}4$ supersymmetry acts
on the fields $f^{iA}$ and $\chi^{aA}\,$. An equivalent realization
on the complex fields defined in \p{redef1}, \p{redef2} is presented
in Appendix A.

The realization of $\mathcal{N}{=}4$ supersymmetry in the standard $\mathcal{N}{=}4$ superspace $(t, \theta^{ia})$
is given by \p{st}.
Then the harmonic projections of $\theta^{ia}$, i.e. $\theta^{m,n} \equiv \theta^{ia}u^m_i v^n_a\,, \, m{=}\pm 1\,, n{=}\pm 1\,,$
are transformed as
\be
\delta \theta^{m, n} = \epsilon^{m, n}\,,
\ee
while the ``analytic'' time coordinates $t_{\pm}$ defined in \p{anb1} and \p{anb2} as
\be
\delta t_+ = 2i(\epsilon^{-1,1}\theta^{1, -1} - \epsilon^{-1,-1}\theta^{1, 1})\,, \quad
\delta t_- = 2i(\epsilon^{1,-1}\theta^{-1, 1} - \epsilon^{-1,-1}\theta^{1, 1})\,,
\ee
thus confirming that the analytic subspaces \p{anss1} and \p{anss2} are closed under $\mathcal{N}{=}4$ supersymmetry.
Using these coordinate transformations, it is straightforward to find the transformation laws of the component fields in
the analytic superfields $q^{1,0}(\zeta_+,u,v)$ and $\psi^{0,1}(\zeta_-,u,v)$ defined by
the $\theta$-expansions \p{sq} and \p{spsi}:
\be
\delta f^{iA} = \epsilon^{ia}\psi_{a}^{A}, \quad \delta\psi^{aA} = -2i\epsilon^{ia}\dot{f}_{i}^{A}\,,
\ee
and
\be
\delta\chi^{aA} = \epsilon^{ia}h_{i}^{A}, \quad \delta h^{iA} = -2i\epsilon^{ia}\dot{\chi}_{a}^{A}.
\ee

In order to find the supersymmetry  transformations in terms of the physical fields only, we should express
the auxiliary fields $h^{iA}$ and $\psi^{aA}$ from their equations of motion \p{h} and \p{psi}.
As a result, we obtain
\be
\delta f^{iA} = -\frac{i}{\sqrt{\kappa}}\,\epsilon^{ia}C^{AB}\dot{\chi}_{aB}, \quad
\delta\chi^{aA} = -\frac{1}{\sqrt{\kappa}}\,\epsilon^{ia}\dot{f}_{i}^{A}. \lb{tranS}
\ee
The variation of the Lagrangian \p{Lfchi} under these transformations is equal to
\be
\delta L = i\sqrt{\kappa}\,\epsilon^{ia}\partial_{t}\left(\chi_{aA}\dot{f}_{i}^{A} -\dot{\chi}_{aA}f_{i}^{A} -
- \frac{1}{\kappa}C^{AB}\dot{\chi}_{aA}\dot{f}_{iB}\right).
\ee
The corresponding conserved Noether supercharge is defined in the standard way
\be
\epsilon^{ia}S_{ia} = \delta f^{iA}\frac{\partial L}{\partial\dot{f}^{iA}}
+ \delta \chi^{aA}\frac{\partial L}{\partial\chi^{aA}}
- i\sqrt{\kappa}\,\epsilon^{ia}\partial_{t}\left(\chi_{aA}\dot{f}_{i}^{A}
-\dot{\chi}_{aA}f_{i}^{A} - \frac{1}{\kappa}C^{AB}\dot{\chi}_{aA}\dot{f}_{iB}\right), \nonumber
\ee
and is calculated to be
\be
S_{ia} = -\frac{i}{\sqrt{\kappa}}\,C^{AB}\dot{\chi}_{aA}\dot{f}_{iB}\,.\lb{nc}
\ee
Using equations of motion for the physical fields,
\be
\ddot{f}_{iA} = -2\kappa\,C_{AB}\dot{f}_{i}^{B}, \quad \ddot{\chi}_{a}^{A}
= 2\kappa\,C^{AB}\dot{\chi}_{aB}\,, \lb{eqm}
\ee
it is easy to directly check that $\dot{S}_{ia} = 0\,$.

The Noether charges  \p{nc} become the generators of $\mathcal{N}{=}4$ supersymmetry upon quantization.

\subsection{Target space supersymmetry}
Before turning to quantization of $\mathcal{N}{=}4$ supersymmetric
Landau model, let us show  that, besides the worldline
$\mathcal{N}{=}4$ supersymmetry, the model \p{Lfchi} possesses
invariance under certain target space supersymmetry which
generalizes the $ISU(1|1)$ supersymmetry of the ${\cal N}{=}2$ Landau
model. Anticipating the quantum picture, we shall present a
realization of this supersymmetry by differential operators acting
in the target (4 +4) superspace $(f^{iA}, \chi^{aB})\,$. All
these operators are obtained in the standard way from the conserved
Noether charges associated with the appropriate invariances of the
action corresponding to the Lagrangian \p{Lfchi}.

The most evident type of such a symmetry is the ``magnetic'' supertranslations:
\be
\delta f^{iA} = b^{iA}, \quad \delta \chi^{aB} = \nu^{aB},\lb{PPi}
\ee
where $b^{iA}$ and $\nu^{aB}$ are constant bosonic and fermionic parameters.
The corresponding symmetry generators are
\be
P_{iA} = -i\partial_{f^{iA}} + \kappa\, C_{AB}f_i^B, \quad \Pi_{aA} = \partial_{\chi^{aA}} + \kappa\, \chi_{aA}.\lb{tran}
\ee

There are also two automorphism groups $SU(2)_L$ and $SU(2)_R\,$, which separately rotate the indices $i$ and $a$ and
so are realized, respectively,  on the bosonic and fermionic fields:
\be
\delta f^{iA} = \lambda^i_j f^{jA}, \quad \delta \chi^{aA} = \lambda^a_b \chi^{bA},\lb{TbTf}
\ee
with $\lambda_{ij} = \lambda_{ji}$ and $\lambda_{ab} = \lambda_{ba}$. The corresponding generators are
\be
 T^{(i}_{\quad j)} = f^{iA}\partial_{f^{jA}} - \frac{1}{2}\delta^i_jf^{k A}\partial_{f^{k A}},
\quad T^{(a}_{\quad b)} = \chi^{aA}\partial_{\chi^{bA}} - \frac{1}{2}\delta^a_b\chi^{b A}\partial_{\chi^{b A}}.\lb{rot}
\ee

There is also $U(1)$ symmetry which simultaneously changes the phase of all fields:
\be
\delta f^{iA} = \alpha\, C^A_B\,f^{iB}, \quad \delta \chi^{aA} = \alpha\, C^A_B\,\chi^{aB}.\lb{I1}
\ee
The corresponding generator is \footnote{In addition to \p{I1}, one can define a similar independent $U(1)$ symmetry
rotating only fermions, i.e.
$\delta \chi^{aA} = \beta\, C^A_B\,\chi^{aB}\,, \;\delta f^{iA} = 0\,$. This additional symmetry can be treated
as some automorphism of the
full target space symmetry. Taking it into account, the Lagrangian \p{Lfchi} exhibits
four independent $U(1)$ invariances,
which become manifest in the complex notation \p{Lz}.}
\be
Z = -iC^A_B(f^{iB}\partial_{f^{iA}} + \chi^{aB}\partial_{\chi^{aA}}).\lb{I}
\ee

Finally, there are odd linear symmetries which mix $f^{iA}$ with $\chi^{aA}$:
\bea
\delta f^{iA} = \frac{1}{2}\omega^{ia}(C^A_B + i\delta^A_B)\chi^B_a
- \frac{1}{2}\bar{\omega}^{ia}(C^A_B - i\delta^A_B)\chi^B_a\,, \nonumber \\
\delta\chi^{aA} = \frac{1}{2}\omega^{ia}(C^A_B - i\delta^A_B)f^B_i
+ \frac{1}{2}\bar{\omega}^{ia}(C^A_B + i\delta^A_B)f^B_i\,.\lb{oddQ1}
\eea
The relevant generators are
\bea
Q^{ia} = \frac{1}{2}(iC^A_B - \delta^A_B)\chi^{aB}\partial_{f^A_i}
+ \frac{1}{2}(iC^A_B + \delta^A_B)f^{iB}\partial_{\chi^A_a}\,, \nonumber \\
\bar{Q}_{ia} = \frac{1}{2}(iC^A_B + \delta^A_B)\chi^B_a\partial_{f^{iA}}
-\frac{1}{2}(iC^A_B - \delta^A_B)f^B_i\partial_{\chi^{aA}}\,.\lb{oddQ}
\eea

Having the explicit form of the generators, it is easy to establish
the algebra of their (anti)commutators.

The generators \p{tran} form a superalgebra of magnetic supertranslations
\be
[P_{iA}, P_{jB}] = 2i \kappa\epsilon_{ij} C_{AB}\,, \quad
\{\Pi_{aA}, \Pi_{bB}\} = 2\kappa\epsilon_{ab}\epsilon_{AB}\,. \lb{PPi1}
\ee
The generators \p{rot}, \p{I}, \p{oddQ} form the superalgebra $su(2|2)$:
\be
\{Q^{ia}, \bar{Q}_{jb}\} = \delta^a_b{T}^{(i}_{\quad j)} - \delta^i_jT^{(a}_{\quad b)}
- \frac{1}{2}\delta^i_j\delta^a_b\,Z, \lb{QQ}
\ee
\be
\{Q^{ia}, Q^{jb}\} = \{\bar{Q}_{ia}, \bar{Q}_{jb}\} = 0, \lb{QQ1}
\ee
\be
[Q^{ia}, Z] = 0, \quad [\bar{Q}_{jb}, Z] = 0, \lb{QI}
\ee
\be
[\bar{Q}_{ia}, {T}^{(j}_{\quad k)}] = \delta^j_i\bar{Q}_{ka}
- \frac{1}{2}\delta^j_k\bar{Q}_{ia}, \quad [\bar{Q}_{ia}, T^{(b}_{\quad c)}] = \delta^b_a\bar{Q}_{ic}
- \frac{1}{2}\delta^b_c\bar{Q}_{ia}, \lb{QT1}
\ee
\be
[Q^{ia}, {T}^{(j}_{\quad k)}] = -\delta^i_k Q^{ja} + \frac{1}{2}\delta^j_kQ^{ia}, \quad [Q^{ia}, T^{(b}_{\quad c)}]
= -\delta^a_c Q^{ib} + \frac{1}{2}\delta^b_cQ^{ia}, \lb{QT2}
\ee
\be
[Z, {T}^{(i}_{\quad j)}] = [Z, T^{(a}_{\quad b)}] = 0, \lb{IT1}
\ee
\be
[{T}^{(i}_{\quad j)}, {T}^{(k}_{\quad l)}] = \delta^k_j {T}^{(i}_{\quad l)}
- \delta^i_l{T}^{(k}_{\quad j)}, \quad [T^{(a}_{\quad b)}, T^{(c}_{\quad d)}] = \delta^c_bT^{(a}_{\quad d)}
- \delta^a_d T^{(c}_{\quad b)}. \lb{TT1}
\ee
We employ the following rules of hermitian conjugation: $(P^{iA})^\dag = P_{iA}\,,\, (\Pi^{iA})^\dag = \Pi_{iA}\,$,
$(Q^{ia})^\dag = -\bar{Q}_{ia}\,,\, (T^{(ij)})^\dag = - T_{(ij)}\,,\, (T^{(ab)})^\dag = - T_{(ab)}\,, \,Z^\dag = Z\,$.
Note that the generator $Z$ is the ``central charge'' generator. It has a non-trivial
realization on the fields $(f^{iA}, \chi^{aB})$
(see \p{I}), so in the present case we cannot factor it out to end up with the supergroup $PSU(2|2)\,$.

Finally, we present the commutation relations between the generators of the magnetic supertranslation group
and the $SU(2|2)$ generators
\be
[Z, P_{iA}] = iC^B_A P_{iB}, \quad [Z, \Pi_{aA}] = iC^B_A \Pi_{aB}, \lb{PI}
\ee
\be
[P_{iA}, {T}^{(j}_{\quad k)}] = \delta^j_iP_{kA} - \frac{1}{2}\delta^j_kP_{iA}, \quad [\Pi_{aA}, T^{(b}_{\quad c)}]
= \delta^b_a\Pi_{cA} - \frac{1}{2}\delta^b_c\Pi_{aA}, \lb{TPPi}
\ee
\be
[Q^{ia}, P_{jA}] = -\frac{i}{2}\delta^i_j\epsilon^{ab}(iC_A^B + \delta_A^B)\Pi_{bB}, \quad \{Q^{ia}, \Pi_{bA}\}
= -\frac{i}{2}\delta^a_b\epsilon^{ij}(iC_A^B - \delta_A^B)P_{jB}, \lb{QP}
\ee
\be
[\bar{Q}_{ia}, P_{jA}] = -\frac{i}{2}\epsilon_{ij}(iC_A^B - \delta_A^B)\Pi_{aB}, \quad
\{\bar{Q}_{ia}, \Pi_{bA}\} = \frac{i}{2}\epsilon_{ab}(i C_A^B + \delta_A^B)P_{iB}. \lb{QPi}
\ee

Thus the algebra of the  magnetic (super)translation generators forms an ideal
in the full target supersymmetry algebra,
which is the semi-direct product
\be
(P_{iA}, \Pi_{aA})\rtimes SU(2|2) = ISU(2|2).
\ee
Correspondingly, the $(4|4)$-dimensional target manifold of the physical
fields $f^{iA}, \chi^{aB}$ can be identified with
the supercoset $ISU(2|2)/SU(2|2)\,$.

Recall that our original propositions were the requirement of manifest ${\cal N}{=}4$ worldline supersymmetry
and a sort of minimality principle: we wished to construct a model which would be a minimal generalization of ${\cal N}{=}2$
Landau model. And finally we found that the model constructed possesses, as a gift, an extra target supersymmetry $ISU(2|2)$!
As is shown in Appendix B, this supergroup also admits an off-shell realization on the bi-harmonic superfields $q^{1,0 A}$
and $\psi^{0, 1 A}\,$.

It should be also pointed out that the extended target $SU(2|2)$ supersymmetry is just an extension of the target $SU(1|1)$
supersymmetry of the ${\cal N}{=}2$ Landau model \cite{3,32,33}. The Lagrangian \p{Lz} possesses two mutually commuting $SU(1|1)$
symmetries of this type realized on the pairs of $d{=}1$ fields $(z,\zeta)$ and $(u, \xi)\,$. In addition, it is invariant under two
extra $SU(1|1)$ symmetries realized on $(z, \xi)$ and $(u, \zeta)\,$, which commute with each other, but not with the previous
two $SU(1|1)$. The full symmetry $SU(2|2)$ is none other than the minimal closure of these different
$SU(1|1)$ symmetries of \p{Lz}. The explicit realization of their generators is given in Appendix A. The fields
$(z,  u, \xi, \zeta)$ transform according to a fundamental representation of $SU(2|2)$. We also note that the appearance
of this extended target space supersymmetry is of course a consequence of the worldline ${\cal N}{=}4$ supersymmetry because the latter requires
additional bosonic and fermionic fields (as compared to the field contents of the ${\cal N}{=}2$ model) for arranging
irreducible ${\cal N}{=}4$ supermultiplets. However, it essentially relies as well upon our particular choice of
the superfield action \p{supact} which directly generalizes the ${\cal N}{=}2$ action \p{8}. In Section 7 we will
consider a more general superfield action that is still ${\cal N}{=}4$ supersymmetric (by construction), but possesses
no target space $ISU(2|2)$ supersymmetry.

\section{Quantization}
\setcounter{equation}{0}
It is convenient to perform quantization in terms of the complex fields $z, u, \zeta, \xi$, so
in this Section we will proceed from the Lagrangian \p{Lz}.

\subsection{Hamiltonian}
The canonical momenta for the bosonic and fermionic  fields defined as $\pi_b = \frac{\partial L}{\partial \dot b}$ and
 $\pi_f = \frac{\partial L}{\partial \dot f}$ are given by
\bea
\pi_{z} = \dot{\bar{z}} - i\kappa\bar{z}, \quad \pi_{\bar{z}} = \dot{z} + i\kappa z, \nonumber \\
\pi_{u} = \dot{\bar{u}} - i\kappa\bar{u}, \quad \pi_{\bar{u}} = \dot{u} + i\kappa u\,,
\eea
and
\bea
\pi_{\zeta} = \dot{\bar{\zeta}} -i \kappa\bar{\zeta}, \quad \pi_{\bar{\zeta}} = -\dot{\zeta} -i \kappa\zeta, \nonumber \\
\pi_{\xi} = \dot{\bar{\xi}} -i \kappa\bar{\xi}, \quad \pi_{\bar{\xi}} = -\dot{\xi} -i \kappa\xi.
\eea
The classical Hamiltonian is
\be
H_{cl} = (\pi_{z} + i\kappa \bar{z})(\pi_{\bar{z}} - i\kappa z) + (\pi_{u} + i\kappa\bar{u})(\pi_{\bar{u}} - i\kappa u)
- (\pi_{\bar{\zeta}} +i \kappa\zeta)(\pi_{\zeta} +i \kappa\bar{\zeta}) -(\pi_{\bar{\xi}} + i\kappa\xi)(\pi_{\xi}
+i \kappa\bar{\xi}).\lb{clash}
\ee
We quantize by the substitution
\be
\pi_{b} \rightarrow -i\frac{\partial}{\partial b}, \quad \pi_{f}   \rightarrow -i \frac{\partial}{\partial f},
\ee
and define the quantum Hamiltonian as the Weyl-ordered form of \p{clash}
\be
H_q = a_{z}^{\dag}a_{z} + a_{u}^{\dag}a_{u} - \alpha_{\zeta}^{\dag}\alpha_{\zeta} - \alpha_{\xi}^{\dag}\alpha_{\xi}\,.
\ee
Here,
\be
a_{z}^{\dag} = i(\frac{\partial}{\partial z} - \kappa \bar{z}), \quad a_{z}
= i(\frac{\partial}{\partial \bar{z}} + \kappa z)\,, \quad a_{u}^{\dag} = i(\frac{\partial}{\partial u}
- \kappa \bar{u}), \quad a_{u}
= i(\frac{\partial}{\partial \bar{u}} + \kappa u)\,, \nonumber
\ee
\be
[a_{z}, a_{z}^{\dag}] = 2\kappa\,, \quad [a_{u}, a_{u}^{\dag}] = 2\kappa\,,\lb{a-com}
\ee
and
\be
\alpha_{\zeta}^{\dag} = \frac{\partial}{\partial\zeta} - \kappa\bar{\zeta}\,, \quad \alpha_{\zeta} =
\frac{\partial}{\partial\bar{\zeta}} - \kappa\zeta\,, \quad \alpha_{\xi}^{\dag}
= \frac{\partial}{\partial\xi} - \kappa\bar{\xi}\,, \quad \alpha_{\xi} = \frac{\partial}{\partial\bar{\xi}}
- \kappa\xi\,, \nonumber
\ee
\be
\{\alpha_{\zeta}, \alpha_{\zeta}^{\dag}\} = -2\kappa\,, \quad \{\alpha_{\xi}, \alpha_{\xi}^{\dag}\}
= -2\kappa\,. \lb{alpha-com}
\ee

Note that, like in the ${\cal N}{=}2$ Landau model \cite{3,33}, the Hamiltonian admits
a nice Sugawara-type representation
\be
H_q = \frac{1}{2}P^{iA}P_{iA} + \frac{i}{2}C^{AB}\Pi^{a}_{A}\Pi_{aB} - 2\kappa Z\,, \lb{SugH}
\ee
which means that it belongs to the enveloping algebra of the superalgebra $ISU(2|2)$ defined in the previous Section.
Using this form of
$H_q\,$, it is straightforward to check that it commutes with all $ISU(2|2)$ generators.

Sometimes it is useful to know the explicit form of the Hamiltonian
\be
H_q = -\Big(\frac{\partial}{\partial z}\frac{\partial}{\partial \bar z} + \frac{\partial}{\partial u}
\frac{\partial}{\partial \bar u}
+ \frac{\partial}{\partial \zeta}\frac{\partial}{\partial \bar \zeta} + \frac{\partial}{\partial \xi}
\frac{\partial}{\partial \bar \xi}\Big)
+\kappa^2\,(|z|^2 + |u|^2 + \zeta\bar \zeta + \xi\bar \xi) - \kappa\,Z\,, \lb{Hexpl}
\ee
where, in the complex notation, the U(1) generator $Z$ defined in \p{I} is expressed as
\be
Z = z \frac{\partial}{\partial z} - \bar z \frac{\partial}{\partial \bar z} +
u \frac{\partial}{\partial u} - \bar u \frac{\partial}{\partial \bar u} + \zeta \frac{\partial}{\partial \zeta}
- \bar \zeta \frac{\partial}{\partial \bar \zeta} + \xi \frac{\partial}{\partial \xi}
- \bar \xi \frac{\partial}{\partial \bar \xi}\,.
\ee

\subsection{Wave functions and degeneracies}
\noindent{\bf LLL.} By definition, the wave function of the Lowest Landau Level (LLL) $\Psi^0$ is nullified
by both bosonic and fermionic
annihilation operators $a_z\,, a_u\,, \alpha_{\zeta}$ and $\alpha_{\xi}$:
\bea
(\partial_{\bar{z}} + \kappa z)\Psi^0 = (\partial_{\bar{u}} + \kappa u)\Psi^0
= (\partial_{\bar{\zeta}} - \kappa\zeta)\Psi^0 = (\partial_{\bar{\xi}} - \kappa\xi)\Psi^0 = 0
\Leftrightarrow \nonumber H_q\,\Psi^0 = 0\,.
\eea
These conditions can be solved in terms of the holomorphic ``reduced'' wave function $\psi_0$:
\bea
\Psi^0 = e^{-\kappa K}\psi_{0}(z,u,\zeta,\xi), \quad K = |z|^2 + |u|^2 + \zeta\bar{\zeta} + \xi\bar{\xi}\,.
\eea
The LLL has a four-fold degeneracy,
\be
\psi_{0}(z,u,\zeta,\xi) = A^{0}(z,u) + \zeta B^{0}(z,u) + \xi C^{0}(z,u) + \zeta\xi D^{0}(z,u)\,,
\ee
where $A^0, B^0, C^0, D^0$ are analytic functions of $z$ and $u\,$. Their set is closed under the action of $ISU(2|2)$
\footnote{Each of these four LLL states is also {\it infinitely} degenerated due to the symmetry under
the ``magnetic'' translations. This degeneracy generalizes the similar phenomenon in the standard
bosonic $2D$ Landau model and
its ${\cal N}{=}2$ extension.}.
\vspace{0.2cm}

\noindent{\bf Excited LLs.} The Hilbert space for the $N$-th Landau level is spanned by the wave functions:

\be
\Psi^{(N)} \sim \sum_{j=0}^{N}(a_z^{\dag})^{j}(a_u^{\dag})^{N-j}
e^{-\kappa K}\psi_{(0,0)}^{(j,N-j)}(z,u,\zeta,\xi) \nonumber
\ee
\be
+ \sum_{j=0}^{N-1}(a_z^{\dag})^{j}(a_u^{\dag})^{N-1-j}\Big[\alpha_{\zeta}^{\dag}e^{- \kappa K}
\psi_{(1,0)}^{(j,N-1-j)}(z,u,\zeta,\xi)
+ \alpha_{\xi}^{\dag}e^{-\kappa K}\psi_{(0,1)}^{(j,N-1-j)}(z,u,\zeta,\xi)\Big]  \nonumber
\ee
\be
+\,\sum_{j=0}^{N-2}(a_z^{\dag})^{j}(a_u^{\dag})^{N-2-j}\alpha_{\zeta}^{\dag}\alpha_{\xi}^{\dag}
e^{-\kappa K}\psi_{(1,1)}^{(j,N-2-j)(z,u,\zeta,\xi)}(z,u,\zeta,\xi)\,, \lb{wf1}
\ee
\be
H_q\Psi^{(N)} = 2\kappa N\,\Psi^{(N)}\,,\lb{wf2}
\ee
where
\bea
\psi_{(l,m)}^{(j,N-j-l-m)}(z,u,\zeta,\xi) = A_{(l,m)}^{(j,N-j-l-m)} + \zeta B_{(l,m)}^{(j,N-j-l-m)}
+ \xi C_{(l,m)}^{(j,N-j-l-m)}
+ \zeta\xi D_{(l,m)}^{(j,N-j-l-m)}\nonumber \\
\eea
and the indices $l,m = 0,1\,$ represent the numbers of fermionic excitations produced
by $\alpha^\dagger_\zeta$ and
$\alpha^\dagger_\xi$. One can rewrite \p{wf1} in another way,
\be
\Psi^{(N)} = \sum_{j=0}^{N}\Psi^{(j,N-j)}_{(0,0)} + \sum_{j=0}^{N-1}\Psi^{(j,N-1-j)}_{(1,0)}
+ \sum_{j=0}^{N-1}\Psi^{(j,N-1-j)}_{(0,1)}
+ \sum_{j=0}^{N-2}\Psi^{(j,N-2-j)}_{(1,1)}\,,
\ee
where
\be
\Psi^{(j,N-j-l-m)}_{(l,m)} = (a_z^{\dag})^{j}(a_u^{\dag})^{N-j-l-m}(\alpha_{\zeta}^{\dag})^{l}(\alpha_{\xi}^{\dag})^{m}
e^{-\kappa K}\psi_{(l,m)}^{(j,N-j-l-m)}(z,u,\zeta,\xi)\,.
\ee
This state describes the system with energy $2\kappa N$ and with $j$ and $N-j-l-m$ excited quanta
of the bosonic fields $z$ and $u\,$,
respectively, and with $l$ and $m$ excited quanta of the fermionic fields $\zeta$ and $\xi\,$, respectively.
LL with $N>0$ has a degeneracy $4(N+1) + 4N + 4N + 4(N-1) = 16N\,$ (modulo an infinite degeneracy due
to the invariance under bosonic magnetic translations).

In order to better understand the origin of this degeneracy of the $N$-th LL, it is convenient
to pass to the $SU(2)_{L, R}$
covariant notation for the creation and annihilation operators, $a^i := (a_z,\;a_u)\,, \;a^{\dag}_i :=
(a_{\bar z}, \;a_{\bar u})\,, \;\alpha^a := (\alpha_\zeta,\;\alpha_\xi)\,, \;\alpha^{\dag}_a := (\alpha_{\bar\zeta}, \;
\alpha_{\bar\xi})\,$,
\be
[a^{\dag}_i, a^j] = -2\kappa\delta^j_i, \quad \{\alpha^{\dag}_a, \alpha^b\} = -2\kappa\delta_a^b\,.
\ee
Then the Hamiltonian can be rewritten as
\be
H_q = a^{\dag}_ia^i - \alpha^{\dag}_a\alpha^a\,.
\ee
With this notation, the wave functions are
\bea
\Psi^{(N)} &=& a^{\dag}_{(i_1}a^{\dag}_{i_2}...a^{\dag}_{i_N)}e^{-\kappa K}\phi^{(i_1i_2...i_N)}(z,u,\zeta,\xi) +
\alpha^{\dag}_aa^{\dag}_{(i_1}a^{\dag}_{i_2}...a^{\dag}_{i_{N-1)}}
e^{-\kappa K}\psi^{a(i_1i_2...i_{N-1})}(z,u,\zeta,\xi) \nonumber \\
&& +\, (\alpha^{\dag})^2a^{\dag}_{(i_1}a^{\dag}_{i_2}...a^{\dag}_{i_{N-2)}}
e^{-\kappa K}\phi^{(i_1i_2...i_{N-2})}(z,u,\zeta,\xi).\lb{Deg}
\eea
Now it becomes obvious why the degeneracy of the $N$-th LL is just $16N$. The component
wave functions  $\phi^{(i_1i_2...i_N)}\,$,
$\psi^{a(i_1i_2...i_{N-1})}$ and $\phi^{(i_1i_2...i_{N-2})}$ in \p{Deg} are irreducible tensors of $SU(2)_L$ with
the spins $s_1 = \frac{N}{2}$, $s_2 = \frac{N-1}{2}$ (entering twice) and $s_3 = \frac{N-2}{2}\,$, respectively.
So the degeneracy of the $N$-th level is equal to
\be
4[(2s_1 + 1) + 2(2s_2 + 1) + (2s_3 + 1)] = 16N\,.
\ee

The wave function of LLL is a singlet of $\mathcal{N}{=}4$
supersymmetry, while wave functions for any $N>0$ form an
$\mathcal{N}{=}4$ supermultiplet with the $SU(2)_L$ spin contents
$(N/2, \; 2\times(N-1)/2, \; N/2 - 1)$. The number of bosonic
complex fields ($2N$) is always equal to the number of fermionic
complex fields, as it should be.  For example, the first excited level
is described by the wave function \be \Psi^{(N=1)} =
a^{\dag}_{i}e^{-\kappa K}\phi^i + \alpha^{\dag}_a e^{-\kappa
K}\psi^a\,, \ee which corresponds to a ``hypermultiplet'' with the $SU(2)_L$
spin content $(1/2, 0)\,$ (and $(0, 1/2)\,$ with respect
to $SU(2)_R\,$). More details on the realization of ${\cal N}{=}4$ supersymmetry
on the wave functions are given in Sect.~6.2.

\subsection{The problem of negative norms}
As in the case of quantum $\mathcal{N}{=}2$ Landau model, in its $\mathcal{N}{=}4$  extension the wave functions
associated with some levels possess negative norms with respect to the natural $ISU(2|2)$-invariant
inner product defined as
\be
\langle\phi|\psi\rangle = \int d\mu\overline{\phi(z,\bar{z},u,\bar{u},\zeta,\bar{\zeta},\xi,\bar{\xi})}
\psi(z,\bar{z},u,\bar{u},\zeta,\bar{\zeta},\xi,\bar{\xi})\,,
\ee
\be
d\mu = dzd\bar{z}dud\bar{u}d\zeta d\bar{\zeta}d\xi d\bar{\xi}\,. \nonumber
\ee

One can calculate norms of the states $\Psi_{(l,m)}^{(j,k)}$ using the relations \p{a-com} and \p{alpha-com}
\be
\langle\Psi_{(l,m)}^{(j,k)}|\Psi_{(l,m)}^{(j,k)}\rangle = (-1)^{l+m}(2k)^{j+k+l+m}\int d\mu e^{-2\kappa K}
\overline{\psi_{(l,m)}^{(j,k)}(z,u,\zeta,\xi)}\psi_{(l,m)}^{(j,k)}(z,u,\zeta,\xi) \nonumber
\ee
\be
= (-1)^{l+m}(2k)^{j+k+l+m}\Big(\|D_{(l,m)}^{(j,k)}\|^2 + 2k\|B_{(l,m)}^{(j,k)}\|^2 + 2k\|C_{(l,m)}^{(j,k)}\|^2
+ 2k^2\|A_{(l,m)}^{(j,k)}\|^2\Big),\lb{spn}
\ee
where
\be
\|f\|^2 := \int dzd\bar{z}dud\bar{u} e^{-2k|z|^2-2k|u|^2} \overline{f(z,u)}f(z,u)\,.
\ee
We see that the states which include one fermionic creation operator indeed possess negative norms.
To get around this difficulty, one is led to introduce a non-trivial metric on the space of quantum states.
It is natural to redefine the inner product as
\bea
&& \langle\langle\psi|\phi\rangle\rangle := \langle G\psi|\phi\rangle\,, \nonumber \\
&& G\Big(\Psi_{(0,0)}^{(j,k)}
+ \Psi_{(1,0)}^{(j,k)} + \Psi_{(0,1)}^{(j,k)} + \Psi_{(1,1)}^{(j,k)}\Big) = \Psi_{(0,0)}^{(j,k)} - \Psi_{(1,0)}^{(j,k)}
- \Psi_{(0,1)}^{(j,k)} + \Psi_{(1,1)}^{(j,k)}\,,
\eea
where
\be
G = 1 + \frac{\alpha_{\zeta}^{\dag}\alpha_{\zeta}}{\kappa} + \frac{\alpha_{\xi}^{\dag}\alpha_{\xi}}{\kappa}
+ \frac{\alpha_{\zeta}^{\dag}\alpha_{\zeta}\alpha_{\xi}^{\dag}\alpha_{\xi}}{\kappa^2} =
(1 - 2n_{\zeta})(1 - 2n_{\xi})\,, \quad
n_{\zeta, \xi}: = -\frac{\alpha^\dagger_{\zeta, \xi}\,\alpha_{\zeta, \xi}}{2\kappa}\,. \lb{G}
\ee
The metric operator $G$ possesses the standard properties \cite{33,4}
\be
[H_q, G] = 0\,, \quad G^2 = 1\,. \lb{G1}
\ee
With respect to the redefined inner product all norms are positive-definite.

It is worth noting that the rules of hermitian conjugation for those operators which do not commute with $G\,$
are changed. Using the property
\be
\langle\langle\psi|Q\phi\rangle\rangle = \langle G\psi|Q\phi\rangle = \langle Q^{\dag}G\psi|\phi\rangle\,,
\ee
and, on the other hand,
\be
\langle\langle\psi|Q\phi\rangle\rangle = \langle\langle Q^{\ddag}\psi|\phi\rangle\rangle
= \langle GQ^{\ddag}\psi|\phi\rangle\,,
\ee
one finds
\be
Q^{\ddag} = G^{-1}Q^{\dag}G = GQ^{\dag}G = Q^{\dag} + G[Q^{\dag}, G]\,.\lb{ddag}
\ee
The creation and annihilation operators do not commute with $G\,$, so, applying the general formula \p{ddag}, we find
\be
\alpha_{\zeta}^{\ddag} = -\alpha_{\zeta}^{\dag}\,, \quad \alpha_{\xi}^{\ddag} = -\alpha_{\xi}^{\dag}\,,
\ee
whence the manifestly positive-definite form for $H_q$ follows
\be
H_q = a_{z}^{\dag}a_{z} + a_{u}^{\dag}a_{u} + \alpha_{\zeta}^{\ddag}\alpha_{\zeta} + \alpha_{\xi}^{\ddag}\alpha_{\xi}\,.
\ee

To avoid a possible confusion, we would like to point out that the above procedure in our ${\cal N}{=}4$ model (and the analogous one, e.g.,
in ${\cal N}{=}2$ planar super Landau model \cite{33}) makes manifest that the ``would-be'' non-unitarity in these quantum-mechanical models
is in fact fake: it is just related to the ineffectual choice of the inner product in the space of quantum states. With the correct choice,
when the appropriate ``metric'' operator is introduced, all norms (as well as the quantum Hamiltonian) are nicely positive-definite,
the states form a complete set, and no any problem with unitarity arises. It is also worth noting that the redefinition of the inner product
preserves the ${\cal N}{=}4$ supermultiplet structure of the space of states discussed in the previous Subsection. If we would just throw away
the fermionic states with negative norms from the very beginning, this structure would inevitably be broken and we would lose the
${\cal N}{=}4$ supersymmetry (as well as the ${\cal N}{=}2$ one) at the quantum level.

\setcounter{equation}{0}
\section{More on the symmetry structure}

\subsection{Quantum generators of $\mathcal{N}=4$ supersymmetry}
Starting from the classical expression \p{nc} for the supercharges $S^{ia}$, after quantization we find
\be
S^{11} = \frac{i}{\sqrt{\kappa}}(\alpha_{\xi}^{\dag}a_{z} - a_{u}^{\dag}\alpha_{\zeta}), \quad
S^{12} = -\frac{i}{\sqrt{\kappa}}(\alpha_{\zeta}^{\dag}a_{z} + a_{u}^{\dag}\alpha_{\xi}), \nonumber
\ee
\be
S^{21} = \frac{i}{\sqrt{\kappa}}(\alpha_{\xi}^{\dag}a_{u} + a_{z}^{\dag}\alpha_{\zeta}), \quad    S^{22}
= \frac{i}{\sqrt{\kappa}}(a_{z}^{\dag}\alpha_{\xi} - \alpha_{\zeta}^{\dag}a_{u}).
\ee
Using the relations $S^{11\dag} = -S^{22}\,, \;S^{12\dag} = S^{21}\,, \;S^{21\dag} = S^{12\dag}\,, \;S^{22\dag} = -S^{11}\,$,
it is convenient to relabel these generators as
\be
S_1 = S^{21}\,, \;\;S_{1}^{\dag} = S^{12}\,, \;\;S_2 = S^{11}\,, \;\;S_{2}^{\dag} = -S^{22}\,,\lb{Redef22}
\ee
\be
\{S_1, S_2\} = \{S_{1}^{\dag}, S_{2}^{\dag}\} = \{S_1, S_{2}^{\dag}\} = \{S_2, S_{1}^{\dag}\} = 0\,, \nonumber
\ee
\be
\{S_1, S_{1}^{\dag}\} = \{S_2, S_{2}^{\dag}\} = -2H\,.
\ee
The $\mathcal{N}{=}4$ supercharges do not commute with the metric operator \p{G}, so one obtains:
\be
S_1^{\ddag} = -S_1^{\dag}, \quad S_2^{\ddag} = -S_2^{\dag}, \lb{dagS}
\ee
\be
\{S_1, S_1^{\ddag}\} = \{S_2, S_2^{\ddag}\} = 2H.
\ee

In the covariant notation, $H$ and $S^{ia}$ are expressed as
\be
S^{ia} = -\frac{i}{\sqrt{\kappa}}(a^i\alpha^{\dag a} - a^{\dag i}\alpha^a), \quad
H = a^{\dag}_ia^i - \alpha^{\dag}_a\alpha^a\,, \lb{Sdag}
\ee
or
\be
S^{ia} = \frac{i}{\sqrt{\kappa}}(a^i\alpha^{\ddag a} + a^{\dag i}\alpha^a), \quad
H = a^{\dag}_ia^i + \alpha^{\ddag}_a\alpha^a,
\ee
and
\be
\{S^{ia}, S^{jb}\} = 2\epsilon^{ij}\epsilon^{ab}H_q\,. \lb{SS}
\ee
The reality properties of $S^{ia}$ with respect to the original and ``improved'' inner products are different,
$$
(S^{ia})^\dag = -S_{ia}\,, \quad (S^{ia})^\ddag = S_{ia}\,,
$$
which agrees with \p{dagS}. It also immediately follows that $S^{ia}$ annihilates the LLL wave function,
i.e. the ground state.
So the ${\cal N}{=}4$ supersymmetry is unbroken in the model under consideration.

More explicit expression for $S^{ia}$ can be found by making, in eq. \p{nc}, the substitutions
$$
\dot{f}_{ia} = \pi_{iA} - \kappa C_{AB}f^B_i \;\Rightarrow \; -(i\partial_{f^{iA}} + \kappa C_{AB}f^B_i)\,,
$$
$$
\dot{\chi}_{aB} = C_B^{\;A}(-i\pi_{aA} + \kappa \chi_{aA}) \; \Rightarrow \; -C_B^{\;A}(\partial_{\chi^{aA}}
- \kappa\chi_{aA})\,.
$$
Using this, one can show that the $\mathcal{N}{=}4$ supercharges, like the Hamiltonian $H_q$ (eq. \p{SugH}),
admit a Sugawara-type
representation in terms of the $ISU(2|2)$ generators,
\be
S^{ia} = 2\sqrt{\kappa}(Q^{ia} + \bar{Q}^{ia}) - \frac{i}{\sqrt{\kappa}}P^i_A\Pi^{aA}, \lb{sugS}
\ee
which implies that they also belong to the enveloping algebra of the $ISU(2|2)$ superalgebra
defined in the previous Section. Calculating the anticommutator of these supercharges with making use of
the (anti)commutation relations of the $ISU(2|2)$ superalgebra alone, one recovers
eq. \p{SS}, with $H_q\,$ given just by the expression \p{SugH}.

\subsection{Second on-shell ${\cal N} = 4$ supersymmetry}
It is rather surprising that, by analogy with the representation \p{sugS}, one can define another set of generators,
\be
\hat{S}{}^{ia} = 2i\sqrt{\kappa}(Q^{ia} - \bar{Q}^{ia}) + \frac{i}{\sqrt{\kappa}}P^i_A\Pi^{a}_BC^{AB}, \quad
(\hat{S}{}^{ia})^\dag = -\hat{S}_{ia}\,, \; (\hat{S}{}^{ia})^\ddag = \hat{S}_{ia}\,,  \lb{sugS2}
\ee
which form the same worldline ${\cal N}{=}4$ superalgebra as $S^{ia}$:
\be
\{\hat{S}{}^{ia}, \hat{S}{}^{jb}\} = 2\epsilon^{ij}\epsilon^{ab}H_q\,. \lb{N42}
\ee
The anticommutator of these two different ${\cal N}{=}4$ supercharges is non-vanishing,
\be
\{S^{ia}, \hat{S}{}^{jb}\} = 8i\kappa\left(\epsilon^{ab}\hat{T}^{ij} - \epsilon^{ij}\hat{T}^{ab} \right),\lb{ShatS}
\ee
where
\bea
&& \hat{T}^{ij} = T^{ij} - \tilde{T}^{ij}\,, \quad \hat{T}^{ab} = T^{ab} - \tilde{T}^{ab}\,, \lb{hatT} \\
&& \tilde{T}^{ij} := -\frac{1}{4i\kappa}\,C^{AB}P^{(i}_A P^{j)}_B\,, \quad \tilde{T}^{ab} :=
-\frac{1}{4\kappa}\,\Pi^{(a}_A \Pi^{b)A}\,. \lb{tildeT}
\eea
The generators $\hat{T}^{ij}, \hat{T}^{ab}$ and $\tilde{T}^{ij}, \tilde{T}^{ab}$ form two mutually commuting sets of
$SU(2)\times SU(2)$
generators. This can be checked using the commutation relations \p{PPi1}, \p{TT1} and \p{TPPi}. Also,
using the important relations
\be
[S^{ia}, P^{i}_A] = [S^{ia}, \Pi^{a}_B] = [\hat{S}{}^{ia}, P^{i}_A] = [\hat{S}{}^{ia}, \Pi^{a}_B] = 0\,, \lb{SPPi}
\ee
which follow from \p{PPi}, one finds that
\be
[S^{ia}, \tilde{T}^{ij}] = [S^{ia}, \tilde{T}^{ab}] = [\hat{S}^{ia}, \tilde{T}^{ij}] =
[\hat{S}^{ia}, \tilde{T}^{ab}] = 0\,,\lb{StildeT}
\ee
and so the $SU(2)$ generators $\hat{T}^{ij}$ and $\hat{T}^{ab}$ act  on $S^{ia}, \hat{S}^{ia}$  in the same way
as the original automorphism $SU(2)_L\times SU(2)_R$ generators $T^{ij}$ and $T^{ab}$. Thus
$\hat{T}^{ij}$ and $\hat{T}^{ab}$ can equally be chosen as generators of the automorphism $SU(2)$ groups of ${\cal N}{=}4$
superalgebras \p{SS} and \p{N42}.

The worldline  superalgebra constituted by the relations \p{SS}, \p{N42} and \p{ShatS} admits a two-fold interpretation.

First, it can be considered as a deformation of the standard ${\cal N}{=}8, d{=}1$ Poincar\'e superalgebra in
the $SO(4)\sim SU(2)_L\times SU(2)_R$ covariant notation (see, e.g., \cite{N8}) by the ``semi-central''
charges $\hat{T}^{ij}$ and $\hat{T}^{ab}$ generating two $SU(2)$ automorphism groups. This deformation makes
the crossing anticommutator $\{S, \hat S\}$ non-vanishing and breaks the $SO(8)$ automorphism group of ${\cal N}{=}8, d{=}1$
superalgebra down to $SO(4)$.

Another interpretation is that the relations \p{SS}, \p{N42} and \p{ShatS} define none other than a second,
``dynamical'' superalgebra $su(2|2)_{dyn}\,$, with the Hamiltonian $H_q$ as the relevant central charge
operator\footnote{We thank Sergey Fedoruk for suggesting
this interpretation to us. It is quite similar to the view of the ${\cal N}{=}2, d{=}1$ Poincar\'e superalgebra
$\{S, \bar S\} = 2H\,, \,S^2 = \bar S^2 =0$ as a sort of $su(1|1)$ superalgebra.}. Indeed, after proper rescaling
of the generators $S^{ia}, \hat S^{ia}, H_q$ and passing to the complex combinations $S^{ia} \pm i \hat S^{ia}$,
the full set of the corresponding (anti)commutation relations can be cast in the standard form \p{QQ} - \p{TT1}.

Finally, we make a few comments on the realization of the hidden worldline $\hat{S}_{ia}$ supersymmetry
on the original fields
$f^{iA}, \chi^{aA}$ and on the wave functions.

The relevant transformations leaving invariant, up to a total derivative,  the on-shell Lagrangian \p{Lfchi} are as follows
\be
\delta f^{iA} = -\frac{i}{\sqrt{\kappa}}{\hat{\epsilon}}{}^{ia}{\dot{\chi}}{}^A_a\,, \quad \delta \chi^{aA} =
-\frac{1}{\sqrt{\kappa}}{\hat{\epsilon}}{}^{ia}C^{AB}\dot{f}_{i B}\,, \lb{tranhat}
\ee
where ${\hat{\epsilon}}{}^{ia}$ is a new quartet Grassmann parameter. The conserved supercurrent, which becomes just \p{sugS2}
after quantization, reads
\be
\hat{S}_{ia} = \frac{i}{\sqrt{\kappa}}\,\dot{\chi}_{aA}\dot{f}{}^{A}_i\,.
\ee
The transformations \p{tranhat}, like \p{tranS},  close on $\partial_t$ only on shell, with taking into account the equations
of motion \p{eqm}. The Lie bracket of \p{tranhat} with \p{tranS} yields an unusual realization of the $SU(2)$ generators
$\hat{T}^{ik}$
and $\hat{T}^{ab}$ on the fields $f^{iA}, \chi^{aA}$: they rotate the latter into their second-order time derivatives
which become the first-order ones only on the shell of eqs.\p{eqm}. These two $su(2)$ algebras are also closed only
modulo \p{eqm}. For the time being, we do not know whether it is possible to reproduce \p{tranhat} from some off-shell
transformations realized on the superfields $q^{1,0 A}, \psi^{0,1 A}\,$.

In the quantum realization via the creation and annihilation operators, the generators $\hat{S}_{1,2}$ related
to $\hat{S}^{ia}$
as in \p{Redef22} are given by the expressions
\be
\hat{S}_1 =\frac{1}{\sqrt{\kappa}}(-a^{\dag}_z\alpha_{\zeta} + a_u\alpha_{\xi}^{\dag}), \quad \hat{S}_2
= \frac{1}{\sqrt{\kappa}}(\alpha^{\dag}_{\xi}a_z + a_u^{\dag}\alpha_{\zeta})\,.
\ee
The corresponding analog of the  $SU(2)_L$ covariant representation \p{Sdag} for $S^{ia}$ reads
\be
\hat{S}^{ia} = -\frac{1}{\sqrt{\kappa}}(a^i\alpha^{\dag a} + a^{\dag i}\alpha^a)\,.\lb{reprSU2}
\ee

It is interesting that the presence of the second worldline ${\cal N}{=}4$ supersymmetry does not give rise
to an additional degeneracy of the wave function: as follows from the representation \p{reprSU2}, action of
$\hat{S}^{ia}$ on the general $N$-th LL  wave function \p{Deg} preserves its structure. In other words, at each LL,
the multiplet of
wave functions closed under the action of $S^{ia}$ is also closed under $\hat{S}^{ia}\,$, and so it carries
an irrep of the whole
worldline supersymmetry $SU(2|2)_{dyn}\,$. This is in striking contrast, e.g., to ${\cal N}{=}8, d{=}1$ supersymmetry
which cannot be realized on a single ${\cal N}{=}4$ multiplet. At least two such multiplets are required
to form ${\cal N}{=}8$ multiplet \cite{N8}. This peculiarity is of course related to the fact that
the anticommutator \p{ShatS} of
$S^{ia}$ and $\hat{S}^{ia}$ is not vanishing: it involves the ``semi-central'' $SU(2)$ generators $\tilde{T}^{ij}$ and
 $\tilde{T}^{ij}$ which have a non-zero action on the wave functions, rotating them with respect to their $SU(2)$
indices. The phenomenon that the presence of central (or ``semi-central'') charges in some supersymmetry algebra
gives rise to ``shortening'' of the relevant irreducible supermultiplets is well known. The case under consideration
supplies one more example of such a situation.

It is instructive to illustrate these features by the example of the wave functions corresponding to the LLs with $N=1$
and $N=2\,$. It is worth noting that the LLL wave function is a singlet of both ${\cal N}{=}4$ supersymmetries
and hence of the entire $SU(2|2)_{dyn}\,$.

The infinitesimal transformations of the general wave function $\Psi^{(N)}$ generated by $S^{ia}$ and $\hat{S}^{ia}$
can be written as
\be
\delta \Psi^{(N)} = -i(\epsilon_{ia}S^{ia} + \hat{\epsilon}_{ia}\hat{S}^{ia})\Psi^{N}\,.\lb{Genvar}
\ee
Using the representations \p{Sdag} and \p{reprSU2} one can explicitly find the transformations induced by \p{Genvar}
for the multiplets of the component wave functions for $N=1$ and $N=2$:
\vspace{0.1cm}
\bea
&& \underline{N=1}: \qquad (\phi^i, \;\psi^a)\,, \nonumber \\
&& \delta \phi^i = -2\sqrt{\kappa}(\epsilon^{ia} + i \hat\epsilon{}^{ia})\,\psi_a, \quad
\delta \psi^a = 2\sqrt{\kappa}(\epsilon^{ia} - i \hat\epsilon{}^{ia})\,\phi_i\,; \lb{N11} \\
&& {\,} \nn
&& \underline{N=2}: \qquad (\phi^{(ij)}, \;\psi^{aj}, \; \phi)\,, \nonumber \\
&&\delta \phi^{(ij)} = -2\sqrt{\kappa}(\epsilon^{(ia} + i\hat\epsilon^{(ia})\,\psi^{j)}_a, \quad \delta \psi^{ai} =
4\sqrt{\kappa}(\epsilon^{ia}+ i\hat\epsilon^{ia})\phi +4\sqrt{\kappa}\,
(\epsilon^{ja}- i \hat\epsilon{}^{ja}) \phi^{(i}_{\;j)}\,, \nn
&& \delta \phi = \sqrt{\kappa}(\epsilon_{ia}- i \hat\epsilon_{ia}) \psi^{ia}\,. \lb{N22}
\eea

One can check that the closure of these transformations agrees with the anticommutation relations  \p{SS}, \p{N42} and \p{ShatS}.
Note that for $N=2$ (and, generally speaking, for all even $N$)  one could formally impose some reality conditions
on the involved functions in a way compatible with either first or second ${\cal N}{=}4$ supersymmetry,
but not with both supersymmetries simultaneously. Thus the component wave functions for any $N$ should
be essentially complex, and
this matches with the property that all of them are holomorphic functions of the complex coordinates
$(z, u, \zeta, \xi)$\footnote{The
irreducible set of wave functions for any $N\geq 1$ is in one-to-one correspondence
(modulo some rescalings) with the $d{=}1$
field content of the complex bi-harmonic analytic
superfield $q^{N,0}(\zeta_+,u,v), D^{2,0}q^{N,0} = D^{0,2}q^{N,0} =0\,$,
in which $i\partial_t$
on all component fields is replaced by $2\kappa N$.}. In fact, the $SU(2|2)_{dyn}$ irreps which are realized
on the ($2N$ + $2N$) multiplets of the $N$-th LL wave functions  are what is called ``atypical''
or ``short'' $SU(2|2)$ representations (see, e.g., \cite{beisert1,beisert,AF, Mi}).

\subsection{Decoupling of worldline and target-space supersymmetries}
Like in the case of ${\cal N}{=}2$ super Landau model \cite{33}, one could ask whether
the above worldline ${\cal N}{=}4$ supersymmetries
are a consequence of the target-space $ISU(2|2)$ symmetry via the Sugawara representation \p{sugS} and \p{sugS2}.
The answer
is that these two types of supersymmetry are in fact independent of each other due to the existence
of the basis in which their
generators are divided into two mutually (anti)commuting sets.

It will be useful to define
\bea
&& \Sigma^{ia}_+ = Q^{ia} + \bar{Q}^{ia}\,, \quad \Sigma^{ia}_- = i(Q^{ia} - \bar{Q}^{ia})\,, \qquad
( \Sigma^{ia}_\pm)^\dag
= -\Sigma^{ia}_\pm\,, \nn
&& \{\Sigma^{ia}_\pm, \Sigma^{jb}_\pm \} = -\epsilon^{ij}\epsilon^{ab} Z\,, \quad \{\Sigma^{ia}_+, \Sigma^{jb}_- \} =
2i\left(\epsilon^{ab}T^{ij} -\epsilon^{ij}T^{ab} \right).\lb{Sigma}
\eea
Then the decoupling transformation is as follows
\be
\tilde{\Sigma}^{ia}_+ = \Sigma^{ia}_+ - \frac{1}{2\sqrt{\kappa}}S^{ia} = \frac{i}{2\kappa}\,P^i_A\Pi^{aA}\,, \quad
\tilde{\Sigma}^{ia}_- = \Sigma^{ia}_- -\frac{1}{2\sqrt{\kappa}}\hat{S}^{ia} = \frac{1}{2i\kappa}\,C^{AB}P^i_A\Pi^{a}_B\,.
\ee
In virtue of the relations \p{SPPi}, $\tilde{\Sigma}^{ia}_\pm$ anticommute with both ${\cal N}{=}4$ supercharges
\be
\{\tilde{\Sigma}^{ia}_\pm, S^{jb} \} = \{\tilde{\Sigma}^{ia}_\pm, \hat{S}^{jb} \} =0\,.
\ee
It is also straightforward to check that $\tilde{\Sigma}^{ia}_\pm$, together with the $SU(2)$ generators
$\tilde{T}^{ij}, \tilde{T}^{ab}$ defined by \p{tildeT}, satisfy just the relations \p{Sigma}, with
\be
\tilde{Z} = Z + \frac{1}{2\kappa}\,H_q\,,
\ee
and have the same (anti)commutation relations \p{QP}, \p{QPi} with the magnetic supertranslation
generators as ${\Sigma}^{ia}_\pm$.
The relations \p{PI}, \p{TPPi}, with $\tilde{Z}, \tilde{T}^{ij}$  and $\tilde{T}^{ab}$ being substituted for
${Z}, {T}^{ij}$ and ${T}^{ab}$, are also satisfied.

Thus, after passing to the generators with tilde, the full symmetry of the ${\cal N}{=}4$ super Landau model
has been reduced to the
direct product $\widetilde{ISU}(2|2)\times SU(2|2)_{dyn}$, with
\bea
\widetilde{ISU}(2|2) \propto \left(P_{iA}, \Pi_{aB}, \tilde{\Sigma}^{ia}_\pm, \tilde{T}^{ij}, \tilde{T}^{ab}, \tilde{Z}\right),
\quad SU(2|2)_{dyn} \propto \left(S^{ia}, \hat{S}^{jb}, \hat{T}^{ij}, \hat{T}^{ab}, H_q \right).
\eea
The generators $\hat{T}^{ij}, \hat{T}^{ab}$ commuting with $\tilde{T}^{ij}, \tilde{T}^{ab}$
(equally as with the remaining generators of $\widetilde{ISU}(2|2)$) were defined in \p{hatT}.
We also took into account the commutation relations \p{StildeT}.

We observe that the $\widetilde{SU}(2|2)$ generators have a Sugawara representation in terms of the supertranslation
generators. On the other hand, if we will try to construct, on the pattern of \p{sugS}, \p{sugS2} and \p{SugH},
some new generators
$S^{ia}, \hat{S}^{ia}$ and $H_q$ out of the $\widetilde{ISU}(2|2)$ generators, they will prove to be identically zero.
Thus in the correctly defined basis, the (anti)commutation relations of the worldline supergroup $SU(2|2)_{dyn}$
do not follow from those of the target-space supergroup $\widetilde{ISU}(2|2)$.

Note that for all generators of $\widetilde{ISU}(2|2)$ the $\ddag$ hermitian conjugation coincides with the ordinary $\dag$
conjugation, whereas
\be
(\Sigma^{ia}_+)^\ddag = (\Sigma^{ia}_+)^\dag - \frac{1}{\sqrt{\kappa}}S^{ia}, \quad (\Sigma^{ia}_-)^\ddag
= (\Sigma^{ia}_-)^\dag - \frac{1}{\sqrt{\kappa}}\hat{S}^{ia}\,.
\ee
Thus the generators of both worldline ${\cal N}{=}4$ supersymmetries appear as the difference between the ``naive'' and ``new''
hermitian conjugates of the spinor generators of the target space $SU(2|2)$ group. In other words, if we would know nothing
about worldline supersymmetries in our model and start, from scratch, with the $ISU(2|2)$ invariant
on-shell component action \p{Lz} as a generalization of \p{12}, we would reveal these supersymmetries as the hidden ones after
passing to the redefined inner product. This is just the way how the worldline ${\cal N}{=}2$ supersymmetry of the planar super Landau model
has been revealed in \cite{33}.

All other features discussed in this Subsection also generalize those found in \cite{33} for the ${\cal N}{=}2$ super Landau model.

\section{Some generalizations}
\setcounter{equation}{0}
In this Section we consider the most general extension of the action  \p{supact} consistent
with the off-shell ${\cal N}{=}4$
supersymmetry, following a similar generalization of the ${\cal N}{=}2$ action \p{8} considered in \cite{4}.

This general ${\cal N}{=}4$ action corresponds to the following modification of first and third terms in \p{supact}:
\bea
S_{gen} &=& \frac{\kappa}{2i}\Big(\int\mu^{-2,0}L^{2,0}(q^{1,0A}, u, v) - i\int\mu^{0,-2}\psi^{0,1A}\psi^{0,1B}
\epsilon_{AB} \nonumber\\
&& +\, \frac{1}{\sqrt{\kappa}}\int\mu^{-2,0}F^{1,0A}(q^{1,0A}, u, v)D^{1,-1}\psi^{0,1B}\epsilon_{AB}\Big), \lb{supactmod}
\eea
where $L^{2,0}$ and $F^{1,0A}$ are arbitrary functions of their arguments. In the presence of non-trivial
function $F^{1,0A}\neq q^{1,0A}$
the equations of motion for auxiliary fields become unsolvable, so in what follows  we choose
$F^{1,0A} = q^{1,0A}$ like in \p{supact}\footnote{A similar restriction was imposed on the ${\cal N}{=}2$
superfield Lagrangian in \cite{4}.}. After integration over Grassmann and harmonic variables, we find
\be
L = \dot{f}^{iA}{\cal A}_{iA} -i\kappa \dot{\chi}^{aA}\chi_{aA} + \frac{i}{2}\,\psi^{aA}\psi_a^B G_{AB}
- \frac{\kappa}{2}\,h^{iA}h_{iA}
- \sqrt{\kappa} (\dot{f}^{iA}h_{iA} + \psi^{aA}\dot{\chi}_{aA}), \lb{genLagr}
\ee
where
\be
{\cal A}_{iA}(f) = -\kappa\left.\int dudv\, u_i^{-1}\frac{\partial L^{2,0}}{\partial q^{1,0A}}\right|_{\theta = 0}, \lb{sdpot}
\ee
\be
G_{AB}(f) = \frac{\kappa}{2}\int dudv\, g_{AB}(f^{iA}u_i^1,u,v), \quad g_{AB}(f^{iA}u_i^1,u,v)
= \left.\frac{\partial^2L^{2,0}}{\partial q^{1,0A}\partial q^{1,0B}}\right|_{\theta = 0}. \lb{Gab}
\ee
Using these definitions, it is easy to show that the background gauge potential ${\cal A}^{iA}$ satisfies
the $\mathbb{R}^4$
self-duality condition
\be
{\cal F}_{iB\, jA} := \partial_{iB} {\cal A}_{jA} - \partial_{jA} {\cal A}_{iB}
= -2G_{AB}\epsilon_{ij}\,.
\ee
The Bianchi identity $\partial_i^{\;A}G_{AB} = 0$ implying that $G_{AB}$ is harmonic,
$\partial^{iC}\partial_{iC}\,G_{AB} = 0\,$,
is automatically satisfied by the expression \p{Gab} for $G_{AB}$, so \p{sdpot} and \p{Gab}
give in fact the most general solution
of the abelian self-duality condition in terms of the analytic harmonic ``prepotential''
$L^{2,0}$ \cite{5}\footnote{The proof can be found in \cite{anal} and \cite{7}.}. Note that the representation
\p{sdpot} also implies
the transversality condition $\partial^{iB}{\cal A}_{iB}= 0\,$, but it can be considered merely
as a gauge choice
because the Lagrangian
\p{genLagr} is invariant, up to total time derivative, under the redefinitions ${\cal A}_{iB} \rightarrow
{\cal A}_{iB} + \partial_{iB}\Lambda(f)\,$.

Thus in the general case the external gauge potential is also self-dual, like in the simplest case \p{supact}.

After eliminating the auxiliary fields, one obtains the following expression for the Lagrangian
in terms of physical fields
\be
L = \dot{f}^{iA} {\cal A}_{iA}- i\kappa\dot{\chi}^{aA}\chi_{aA} + \frac{1}{2}\dot{f}^{iA}\dot{f}_{iA} +
i\,\frac{\kappa}{2}(G^{-1})_{AB}\dot{\chi}^{aA}\dot{\chi}^B_a\,.
\ee
We observe that the bosonic target metric is still flat, in contrast to the general ${\cal N}{=}2$ model of ref. \cite{4}.
The  reason behind this is the impossibility to insert, without breaking of ${\cal N}{=}4$ supersymmetry, any function
of $q^{1,0 A}$ into the second term in \p{supactmod} since this superfield and the fermionic superfield $\psi^{0,1 B}$
live on {\it different} analytic subspaces of the bi-harmonic ${\cal N}{=}4$
superspace. Moreover, it can be shown that any metric can be removed from the kinetic term of $\chi^{a A}$ as well.

To this end, it is convenient to rewrite the Lagrangian in the complex parametrization
\be
L = |\dot{z}|^2 + |\dot{u}|^2 + \dot{z}\bar{\cal A} + \dot{\bar{z}}{\cal A} + \dot{u}\bar{\cal B} + \dot{\bar{u}}{\cal B}
+ iD_{11}\dot{\zeta}\dot{\xi} + iD_{12}\dot{\zeta}\dot{\bar{\zeta}} + iD_{12}\dot{\xi}\dot{\bar{\xi}}
+ iD_{22}\dot{\bar{\zeta}}\dot{\bar{\xi}} - i\kappa(\dot{\zeta}\bar{\zeta} + \dot{\bar{\zeta}}\zeta
+ \dot{\xi}\bar{\xi} + \dot{\bar{\xi}}\xi),
\ee
where ${\cal A}$ and ${\cal B}$ are the components of ${\cal A}^{iA}$ in the complex notation
\be
{\cal A} = {\cal A}^{11}, \quad \bar{\cal A} = {\cal A}^{22}, \quad {\cal B} = {\cal A}^{21}, \quad \bar{\cal B}
= -{\cal A}^{12}\,, \nonumber
\ee
and
\be
D_{AB} = \kappa\,(G^{-1})_{AB}\,. \nonumber
\ee

Under the choice $L^{2,0}(q^{1,0A}, u, v) = q^{1,0A}q^{1,0B}C_{AB}$ we immediately obtain  $D_{AB} = C_{AB}$ and so come back
to the action \p{Lz}.
In the general case, with making use of the parametrization $D_{11} = \bar{D}_{22} = |D_{11}|e^{i\varphi}\,, \;
\overline{(D_{12})} = - D_{12}\,$,  one can find a field redefinition which brings the kinetic terms of the fermion part
into a diagonal form.
This redefinition is as follows
\be
\zeta = \frac{-ie^{-i\varphi}\zeta' + |b|\bar{\xi}'}{\sqrt{1 + |b|^2}}, \quad \xi = \frac{ie^{-i\varphi}\xi'
+ |b|\bar{\zeta}'}{\sqrt{1 + |b|^2}}\,,
\ee
where $|b|$ is sought from the quadratic equation
\be
|b|^2|D_{11}| + 2iD_{12}|b| - |D_{11}| = 0\,. \nonumber
\ee
After some calculation we find the final expression for the action
\be
L = |\dot{z}|^2 + |\dot{u}|^2 + \kappa(\dot{z}\bar{\cal A} + \dot{\bar{z}}{\cal A} + \dot{u}\bar{\cal B}
+ \dot{\bar{u}}{\cal B})
+ \dot{\zeta}'\dot{\bar{\zeta}}' + \dot{\xi}'\dot{\bar{\xi}}' - i\kappa(\dot{\zeta}'\bar{\zeta}'
+ \dot{\bar{\zeta}}'\zeta' + \dot{\xi}'\bar{\xi}' + \dot{\bar{\xi}}'\xi')\,.
\ee

It is rather surprising that the component Lagrangian obtained from \p{supactmod} has the same fermionic part as \p{Lz},
despite the fact that
\p{supactmod} involves the most general interaction. On the other hand, the background gauge field potential consistent
with ${\cal N}{=}4$ supersymmetry turns out to be more general
than just the linear potential (${\cal A} \sim \bar{z}$ and ${\cal B} \sim \bar{u}$) which appears in \p{Lz}
and which was used
in \cite{8} to describe a version of the $\mathbb{R}^4$ QHE. The only constraint on the gauge potential is that
it must be self-dual on $\mathbb{R}^4\,$. It would be interesting to reveal possible physical applications
of such a more general $U(1)$ potential in the QHE on  $\mathbb{R}^4$, e.g.,  along the lines of \cite{8}.

\section{Summary and outlook}
\setcounter{equation}{0}
Let us briefly summarize the results of the paper. We constructed the first example of $\mathcal{N}{=}4$ supersymmetric
Landau model which is a minimal extension of the ${\cal N}{=}2$ super Landau model of ref. \cite{3,33,43}. We started
from the superfield off-shell action involving the linear $\mathcal{N}{=}4$ multiplet $({\bf 4,4,0})$ and
its mirror fermionic counterpart.
After elimination of the auxiliary fields, in the  component Lagrangian there remain four bosonic
and four fermionic physical fields.
In the bosonic limit, when all fermionic fields are suppressed,  one recovers the Lagrangian
of the model used in \cite{8} to describe $\mathbb{R}^4$ quantum Hall effect
with the $U(1)$ background gauge field.
Besides the manifest ${\cal N}{=}4$ supersymmetry, the Lagrangian constructed respects invariance under the target
graded supersymmetry $ISU(2|2)\,$ and, more surprisingly, under the second on-shell ${\cal N}{=}4$ supersymmetry,
which, together with
the first one, close on a ``dynamical'' worldline $SU(2|2)_{dyn}$ symmetry.
We quantized the model and found the spectrum of the Hamiltonian, as well as the degeneracy of
the wave functions for every Landau level.
The LLL wave function is a singlet of ${\cal N}{=}4$ supersymmetries, while the wave functions of the next LLs form
irreducible ${\cal N}{=}4$ (and $SU(2|2)_{dyn}$) multiplets. For the wave functions to possess non-negative
norms and, hence,
for the model to preserve unitarity, one is led, like in the ${\cal N}{=}2$ case, to introduce a non-trivial metric operator
on the space of states and thus to redefine the corresponding inner product. We also discussed
the most general form of the action of the original two multiplets, such that it is compatible with the worldline
$\mathcal{N}{=}4$ supersymmetry.
The latter requirement proves to be very restrictive: as opposed to the generic ${\cal N}{=}2$ super Landau action \cite{4}, its
$\mathcal{N}{=}4$ counterpart  involves no non-trivial target superspace metric. The general restriction on the
external gauge field is that it should satisfy the self-duality condition on $\mathbb{R}^4$.

There are few directions in which the present study could be continued. First, it would be interesting
to construct the quantum version
of the generalized ${\cal N}{=}4$ model considered in Sect. 8 and to reveal its possible applications in the $\mathbb{R}^4$
QHE. It is also of obvious interest to extend it in such a way as to gain a non-trivial target
super-metric in the component Lagrangian,
like in \cite{4}, and so to get, in the bosonic sector, a version of Landau model on a curved four-dimensional manifold.
One way to achieve this consists in replacing the linear bosonic $({\bf 4, 4, 0})$ multiplet and, perhaps,
its fermionic mirror
by their nonlinear counterparts along the line of ref. \cite{DI}. Another possibility is to add the mirror
$q^{0, 1 A'}$ and $\psi^{1, 0 A'}$ superfields to the original set $q^{1,0 A}$ and $\psi^{0, 1A}\,$,
thus passing to a model with the eight-dimensional bosonic target space. After such an extension,
one will be able to insert,
in the second term in \p{supactmod}, a  function of $q^{1,0 A}$ (and a function of $q^{0, 1 A'}$
into the new mirror counterpart
of this term),
without any conflict with the bi-harmonic analyticities. These terms can give rise to some non-trivial super-metric
(and superbackground gauge fields) after eliminating auxiliary fields. Also, in such type of super Landau models one
could hope to find out an off-shell
${\cal N}{=}8$ worldline supersymmetry \cite{N8}. It is also tempting to seek for ${\cal N}{=}4$ superextension
of the Landau-type models on $S^4$
with couplings to an external non-abelian $SU(2)$ gauge field \cite{zh,KN,8}. In the conventional ${\cal N}{=}4$
mechanics such couplings
are introduced \cite{IKS} with the help of the so called spin (or isospin) semi-dynamical supermultiplets \cite{FIL}.
One can hope that
the same mechanism works in the case of ${\cal N}{=}4$ super Landau models too. At last, it is worth noting that, besides
${\cal N}{=} 4$ ``hypermultiplets'' $({\bf 4, 4, 0})$ and $({\bf 0, 4, 4})$ utilized here, there are many other
off-shell ${\cal N}{=} 4$ multiplets,  e.g. $({\bf 3, 4, 1})$ and $({\bf 2, 4, 2})\,$.
They, together with their fermionic counterparts, can be used for constructing alternative ${\cal N}{=} 4$ super
Landau models.

One more possible direction of the future work is related to setting up curved analogs of the $(4|4)$
super Landau model constructed here, such
that the latter is reproduced in some contraction limit, like the standard bosonic Landau model \cite{0} is recovered from
the Haldane model \cite{00} after contraction of $SU(2)$ into $E(2)$, the group of motion of the Euclidean plane.
The $(2|2)$ superplane Landau model can be obtained in a similar way from the Landau model on the supersphere
$SU(2|1)/U(1|1) \sim \mathbb{CP}^{(1|1)}$ \cite{SSph}. In the $(4|4)$ case one can expect an analogous relation with
the Landau model on the projective supermanifold $SU(3|2)/U(2|2) \sim \mathbb{CP}^{(2|2)}\,$. It can be regarded
as a superextension of one of the $SU(3)/U(2)$ models considered in ref.~\cite{KN} and could  be closely
related to the integrable $su(3|2)$ spin chain which, in turn, bears an intimate relation to the planar ${\cal N}{=}4$
SYM theory \cite{beisert} and $AdS_5\times S^5$ superstring \cite{STs}. Finally, $SU(2|2)$  already appeared as
a dynamical symmetry acting in the space of quantum states of the super Landau model
on the superflag $SU(2|1)/[U(1)\times U(1)]$ \cite{SSph}, and it would be interesting to clarify possible
links of this realization with that given in the present paper.

\section*{Acknowledgements}
We thank Sergey Fedoruk and Luca Mezincescu for interest in the work and useful comments.
A partial support from the RFBR grants Nr. 09-01-93107 and Nr. 11-02-90445 is acknowledged.

\setcounter{equation}{0}
\appendix

\section{Symmetries in the complex notation}
It is obvious that the symmetry group of the Lagrangian \p{Lz} includes two $ISU(1|1)$ subgroups (on the fields ($z\,, \zeta$)
and ($u\,, \xi$)),
because \p{Lz} is just a sum of two copies of \p{10}. There are two $SU(2)$ subgroups which rotate the fields
($z\,, u$) and ($\zeta, \xi$).
Finally, there are two subgroups $SU(1|1)$ which are realized on the fields ($z\,, \xi$) and ($u\,, \zeta$). Below we list
all generators of these groups.
\vspace{0.2cm}

We start by defining the generators in the complex notation through those in Sect.~4.3:
\bea
&& Q^{11} = -Q^{\dag}_4\,, \quad Q^{12} = -Q^{\dag}_2\,, \quad Q^{21} = Q^{\dag}_1\,, \quad Q^{22} = Q^{\dag}_3\,, \nonumber \\
&& \bar{Q}_{11} = Q_4\,, \quad \bar{Q}_{12} = Q_2\,, \quad \bar{Q}_{21} = -Q_1\,, \quad \bar{Q}_{22} = -Q_3\,, \lb{A1} \\
&& Z = Z_1 + Z_2\,, \lb{A2} \\
SU(2)_L:&& {T}^{(1}_{\quad 1)} = \frac{1}{2}T_{b3}\,, \quad {T}^{(1}_{\quad 2)} = \frac{1}{2}(T_{b1}+T_{b2})\,,
\quad {T}^{(2}_{\quad 1)} = \frac{1}{2}(T_{b1}-T_{b2})\,, \nonumber \\
SU(2)_R:&& T^{(1}_{\quad 1)} = \frac{1}{2}T_{f3}\,, \quad T^{(1}_{\quad 2)} = \frac{1}{2}(T_{f1}+T_{f2})\,, \quad T^{(2}_{\quad 1)}
= \frac{1}{2}(T_{f1}-T_{f2})\,, \lb{A3} \\
&& P_{11} = P_z\,, \quad P_{12} = -P_{\bar{u}}\,, \quad P_{21} = P_u\,, \quad P_{22} = P_{\bar{z}}\,,\nonumber \\
&& \Pi_{11} = \Pi_{\zeta}\,, \quad \Pi_{12} = -\Pi_{\bar{\xi}}\,, \quad \Pi_{21} = \Pi_{\xi}\,, \quad
\Pi_{22} = \Pi_{\bar{\zeta}}\,. \lb{A4}
\eea
Now we give the explicit expressions for these generators.
\vspace{0.2cm}

\noindent 1. $ISU(1|1)$ realized on ($z, \zeta$):
\be
P_{z} = -i(\partial_z + \kappa\bar{z})\,, \quad P_{\bar{z}} = -i(\partial_{\bar{z}} - \kappa z)\,, \quad
\Pi_{\zeta} = \partial_{\zeta} + \kappa\bar{\zeta}\,, \quad \Pi_{\bar{\zeta}}
= \partial_{\bar{\zeta}} + \kappa\zeta\,, \nonumber
\ee
\be
Z_1 = z\partial_z - \bar{z}\partial_{\bar{z}} + \zeta\partial_{\zeta} - \bar{\zeta}\partial_{\bar{\zeta}}\,, \quad
Q_1 = z\partial_{\zeta} - \bar{\zeta}\partial_{\bar{z}}, \quad Q_1^{\dag}
= \bar{z}\partial_{\bar{\zeta}} + \zeta\partial_{z}\,.
\ee
\noindent 2. $ISU(1|1)$ realized on ($u\,, \xi$):
\be
P_{u} = -i(\partial_u + \kappa\bar{u})\,, \quad P_{\bar{u}} = -i(\partial_{\bar{u}} - \kappa u)\,, \quad
\Pi_{\xi} = \partial_{\xi} + \kappa\bar{\xi}\,, \quad \Pi_{\bar{\xi}} = \partial_{\bar{\xi}} + \kappa\xi\,, \nonumber
\ee
\be
Z_2 = u\partial_u - \bar{u}\partial_{\bar{u}} + \xi\partial_{\xi} - \bar{\xi}\partial_{\bar{\xi}}\,, \quad
Q_2 = u\partial_{\xi} - \bar{\xi}\partial_{\bar{u}}\,, \quad Q_2^{\dag} = \bar{u}\partial_{\bar{\xi}} + \xi\partial_{u}\,.
\ee
\noindent 3. $SU(2)_L$ realized on ($z\,, u$):
\be
T_{b1} = z\partial_{u} + u\partial_{z} - \bar{z}\partial_{\bar{u}} - \bar{u}\partial_{\bar{z}}, \;
T_{b2} = z\partial_{u} - u\partial_{z} + \bar{z}\partial_{\bar{u}} - \bar{u}\partial_{\bar{z}},\;
T_{b3} = z\partial_{z} - \bar{z}\partial_{\bar{z}} - u\partial_{u} + \bar{u}\partial_{\bar{u}}\,.
\ee
\noindent 4. $SU(2)_R$ realized on ($\zeta\,, \xi$):
\be
T_{f1} = \zeta\partial_{\xi} + \xi\partial_{\zeta} - \bar{\zeta}\partial_{\bar{\xi}} - \bar{\xi}\partial_{\bar{\zeta}}, \;
T_{f2} = \zeta\partial_{\xi} - \xi\partial_{\zeta} + \bar{\zeta}\partial_{\bar{\xi}} - \bar{\xi}\partial_{\bar{\zeta}}, \;
T_{f3} = \zeta\partial_{\zeta} - \bar{\zeta}\partial_{\bar{\zeta}} - \xi\partial_{\xi} + \bar{\xi}\partial_{\bar{\xi}}\,.
\ee
\noindent 5. Two further $SU(1|1)$ realized on ($z\,, \xi$) and ($u\,, \zeta$):
\bea
Q_3 = z\partial_{\xi} - \bar{\xi}\partial_{\bar{z}}\,, \quad Q_3^{\dag} = \bar{z}\partial_{\bar{\xi}} + \xi\partial_{z}\,,
\quad Z'_1 = z\partial_z - \bar{z}\partial_{\bar{z}} + \xi\partial_{\xi} - \bar{\xi}\partial_{\bar{\xi}} = Z_2 + T_{b3}\,, \\
Q_4 = u\partial_{\zeta} - \bar{\zeta}\partial_{\bar{u}}\,, \quad Q_4^{\dag} = \bar{u}\partial_{\bar{\zeta}} + \zeta\partial_{u}\,,
\quad Z''_2 = u\partial_u - \bar{u}\partial_{\bar{u}} + \zeta\partial_{\zeta} - \bar{\zeta}\partial_{\bar{\zeta}} = Z_1 - T_{b3}\,.
\eea

It is worth noting that only three generators out of the set of $U(1)$ generators $Z_1, Z_2, T_{b3}$, $T_{f3}$
(coming from two supergroups $ISU(1|1)$ and
two groups $SU(2)$) are linearly independent: $Z_2 + T_{b3} + T_{f3} = Z_1\,$. This can be explained as follows.
In the Lagrangian \p{Lfchi}
there are two symmetry automorphism groups $SU(2)_{L,R}$ and an extra group $SU(2)_{ext}$ realized on the indices $A$, but the
latter $SU(2)$ is broken down to some $U(1)$ by constants $C_{AB}$. Thus, there are only three linearly independent mutually
commuting $U(1)$ generators inside $ISU(2|2)$. Note, however, that there is one additional $U(1)$ invariance, which is not contained
in the closure of the odd $ISU(2|2)$ generators. Its generator can be chosen, e.g., as $\zeta\partial_{\zeta}
- \bar{\zeta}\partial_{\bar{\zeta}}\,$. It can be interpreted as an outer automorphism of $ISU(2|2)\,$.
\vspace{0.2cm}

Finally, we rewrite the ${\cal N}{=}4$ supersymmetry transformations \p{tranS} in the complex notation.
For this purpose we introduce complex parameters $\epsilon_1$ and $\epsilon_2\,$,
\be
\epsilon^{11} = -i\epsilon_2\,, \quad \epsilon^{22} = i\bar{\epsilon}_2\,, \quad \epsilon^{21} = i\bar{\epsilon}_1\,,
\quad \epsilon^{12} = i\epsilon_1\,.
\ee
Then the transformations (5.5) take the form
\be
\delta z = \frac{i}{\sqrt{\kappa}}\epsilon_1\dot{\zeta} + \frac{i}{\sqrt{\kappa}}\epsilon_2\dot{\xi}\,,
\quad \delta u = -\frac{i}{\sqrt{\kappa}}\bar{\epsilon}_1\dot{\xi}
+ \frac{i}{\sqrt{\kappa}}\bar{\epsilon}_2\dot{\zeta}\,, \nonumber
\ee
\be
\delta \zeta = \frac{i}{\sqrt{\kappa}}\bar{\epsilon}_1\dot{z} + \frac{i}{\sqrt{\kappa}}\epsilon_2\dot{u}\,, \quad \delta \xi
= \frac{i}{\sqrt{\kappa}}\bar{\epsilon}_2\dot{z} - \frac{i}{\sqrt{\kappa}}\epsilon_1\dot{u}\,.
\ee

\setcounter{equation}{0}
\section{Realization of $ISU(2|2)$ on superfields $q^{1,0A}, \psi^{0,1A}$}
In this appendix we give an off-shell realization of the $ISU(2|2)$ symmetry group on the superfields
$q^{1,0A}, \psi^{0,1A}$,  which in components reproduces the on-shell realization of Sect.~4.3.

We begin with the magnetic supertranslations:
\be
\delta q^{1,0A} = b^{iA}u_i^1, \quad \delta\psi^{0,1A} = \nu^{aA}v_a^1\,. \lb{B1}
\ee

The central charge $Z$ symmetry, which simultaneously changes the phases of all fields, is realized by
\be
\delta q^{1,0A} = \alpha C^A_B q^{1,0B}, \quad \delta\psi^{0,1A} = \alpha C^A_B \psi^{0,1b}\,. \lb{B2}
\ee

The odd $SU(2|2)$ transformations which mix bosonic and fermionic superfields are given by the following variations
\be
\delta q^{1,0A} = D^{1,1}D^{1,-1}\Big(\frac{1}{2}(C^A_B + i\delta^A_B)\Big[A^{-1,-1} + B^{-1,1}D^{0,-2}
+ C^{0,0}D^{-1,-1}\Big]\psi^{0,1B} \nonumber
\ee
\be
-\,\frac{i}{2\sqrt{\kappa}}\frac{1}{2}(C^A_B-i\delta^A_B)\Big[E^{-1,-1}D^{-1,1} + E^{-1,1}D^{-1,-1}
- C^{0,0}D^{-1,-1}D^{-1,1}\Big]q^{1,0B}\Big),\lb{B3}
\ee
\be
\delta \psi^{0,1A} = D^{1,1}D^{-1,1}\Big(\frac{1}{2}(C^A_B - i\delta^A_B)\Big[\hat{A}^{-1,-1} + \hat{B}^{1,-1}D^{-2,0}
- C^{0,0}D^{-1,-1}\Big]q^{1,0B}  \nonumber
\ee
\be
+\,\frac{i}{2\sqrt{\kappa}}\frac{1}{2}(C^A_B+i\delta^A_B)\Big[\hat{E}^{-1,-1}D^{1,-1} + \hat{E}^{1,-1}D^{-1,-1}
+ C^{0,0}D^{-1,-1}D^{1,-1}\Big]\psi^{0,1B}\Big). \lb{SoddQ}
\ee
Here
\be
A^{-1,-1} := \omega^{1,-1}\theta^{-1,1}\theta^{-1,-1}, \quad B^{-1,1} :=\omega^{-1,1}\theta^{1,1}\theta^{-1,-1}
-\omega^{1,1}\theta^{-1,1}\theta^{-1,-1}\,, \nonumber
\ee
\be
C^{0,0} := \omega^{1,-1}\theta^{-1,-1}\theta^{-1,1}\theta^{1,1}-\omega^{1,1}\theta^{-1,-1}\theta^{-1,1}\theta^{1,-1}
+ \omega^{-1,1}\theta^{-1,-1}\theta^{1,1}\theta^{1,-1} - \omega^{-1,-1}\theta^{-1,1}\theta^{1,1}\theta^{1,-1}\,, \nonumber
\ee
\be
E^{-1,-1} := \omega^{-1,1}\theta^{1,-1}\theta^{-1,-1}, \quad E^{-1,1} := \omega^{-1,-1}\theta^{1,1}\theta^{-1,1}\,, \nonumber
\ee
\be
\hat{A}^{-1,-1} := \omega^{-1,1}\theta^{1,-1}\theta^{-1,-1}, \quad \hat{B}^{1,-1} := \omega^{1,-1}\theta^{1,1}\theta^{-1,-1}
-\omega^{1,1}\theta^{1,-1}\theta^{-1,-1}\,, \nonumber
\ee
\be
\hat{E}^{-1,-1} := \omega^{1,-1}\theta^{-1,1}\theta^{-1,-1}\,, \quad \hat{E}^{1,-1}
:= \omega^{-1,-1}\theta^{1,1}\theta^{1,-1}\,. \nonumber
\ee
These superfield transformations amount to the following transformations of the physical and auxiliary fields:
\bea
&& \delta f^{iA} = \frac{1}{2}\omega^{ia}(C^A_B + i\delta^A_B)\chi^B_a\,, \quad \delta \chi^{aA} =
\frac{1}{2}\omega^{ia}(C^A_B - i\delta^A_B)f^B_i\,, \nn
&& \delta h^{ia} = -\frac{1}{2\sqrt{\kappa}}\omega^{ia}(C^A_B + i\delta^A_B)\dot{\chi}^B_a\,, \quad
\delta \psi^{aA} = \frac{1}{2\sqrt{\kappa}}\omega^{ia}(C^A_B - i\delta^A_B)\dot{f}^B_i\,.
\eea
They are consistent with eqs. \p{h}, \p{psi}. The variations with $\bar{\omega}^{ia}$
are obtained from the ${\omega}^{ia}$ ones via the $\sim$ conjugation.

It is straightforward to check the invariance of the superfield action \p{supact} under \p{B1}- \p{SoddQ}. Note that
the structure of the superfield transformations \p{B3} and \p{SoddQ} is almost uniquely determined from the requirement
that their right-hand sides are nullified by the harmonic derivatives $D^{2,0}$ and $D^{0,2}$ (in agreement with the harmonic
constraints (\ref{con_a}b) and (\ref{con_b}b)). All even $SU(2|2)$ transformations (including \p{B2}) are contained in
the closure of \p{B3}, \p{SoddQ} and their $\bar{\omega}^{ia}$ counterparts, so we do not give their explicit form.

\end{document}